\documentclass{aa}  

\usepackage{graphicx}
\usepackage{txfonts}

\usepackage{orcidlink}
\usepackage[switch]{lineno}
\usepackage{booktabs}
\usepackage{caption}
\usepackage{subcaption}
\usepackage{amsmath}
\usepackage{amssymb}
\begin{document}

   \title{Multiwavelength analysis of polarized light in HD~100453
   \thanks{based on observations collected at the European Southern Observatory under ESO programs 110.245U.001 and 105.20JB.001}}
   \author{J. Ma \orcidlink{0000-0003-3583-6652}
          \inst{\ref{inst1}, \ref{inst2}}
          \and R. Tazaki \orcidlink{0000-0003-1451-6836}\inst{\ref{inst3}, \ref{inst1}}
          \and H.M. Schmid\orcidlink{0000-0002-7501-4015}\inst{\ref{inst2}} 
          \and G. Duchêne\orcidlink{0000-0002-5092-6464}\inst{\ref{inst1},\ref{inst4}}
          \and C. Dominik \orcidlink{0000-0002-3393-2459}\inst{\ref{inst5}}
          \and C. Ginski \orcidlink{0000-0002-4438-1971} \inst{\ref{inst6}}
          \and F. Ménard\orcidlink{0000-0002-1637-7393}\inst{\ref{inst1}}
          }
   \institute{Univ. Grenoble Alpes, CNRS, IPAG, F-38000 Grenoble, France \label{inst1}\\
              \email{jie.ma@univ-grenoble-alpes.fr}
              \and {Institute for Particle Physics and Astrophysics, 
              ETH Zurich, Wolfgang Pauli Strasse 17, CH-8093 Zurich, Switzerland} \label{inst2}
              \and {Department of Earth Science and Astronomy, The University of Tokyo, Tokyo 153-8902, Japan} \label{inst3}
              \and {Astronomy Department, University of California Berkeley, Berkeley CA 94720-3411, USA} \label{inst4}
              \and {Anton Pannekoek Institute for Astronomy, University of Amsterdam, Science Park 904, 1098XH Amsterdam, The Netherlands}\label{inst5} 
              \and {School of Natural Sciences, Center for Astronomy, University of Galway, Galway, H91 CF50, Ireland}\label{inst6} 
              }
   \date{Received 30 June 2025; accepted 12 August 2025}
 
  \abstract
   {HD~100453 disk is a prototypical companion-disk interaction system hosting a pair of spirals and a substellar companion.}
   {We present new noncoronagraphic high-contrast imaging observations of HD~100453 with $V$ filter on SPHERE/ZIMPOL. We aim to accurately measure the reflected polarized intensity, the disk intensity, and the degree of polarization at multiwavelength; characterize the dust properties in the disk ring and the companion-driven spiral arms; and search for the radial dependencies of dust scattering properties within the disk.}
   {We combined high-contrast imaging data of the reflected light from 0.55 to 2.2~$\mu m$ using the $V$, $I'$, $J$, and $Ks$ band data of ZIMPOL and IRDIS at VLT/SPHERE.
   For each observational epoch, we corrected for the smearing effect to derive the intrinsic disk-integrated polarized flux. We applied reference differential imaging to extract the disk intensity for the $V$ and $Ks$ bands using star hopping observations. 
   We then used RADMC-3D radiative transfer modeling with a parametrized Henyey-Greenstein phase function to constrain the dust properties.}
   {For the integrated intrinsic polarized flux of the disk, we derived a steady increase with wavelengths from $\hat{Q}_{\varphi}/I_{\star}(V)=0.3\%$ to $\hat{Q}_{\varphi}/I_{\star}(K)=1.2\%$. We obtained the first $V$-band total intensity for HD~100453 with ZIMPOL star hopping. The integrated total flux $I_{\rm disk}/I_{\star}(V)=1.5\%$ increases to $I_{\rm disk}/I_{\star}(K)=4.8\%$. Both the total intensity and the polarization fraction show red colors, and the intrinsic maximum degree of polarization increases moderately from 40\% to 55\%. From the $V$ to $Ks$ band, the dominating dust in the outer disk has an increasing scattering albedo and degree of polarization, while the asymmetry parameter slightly decreases. The outer disk of HD~100453 contains sub-micron-sized low porosity grains/aggregates. The cavity in scattered light is not empty and is replenished with optically thin dust with a maximum size of $\leq 0.1\mu m$. The linear polarization is higher in the spiral region than in the other regions, suggesting different dust properties in those regions.}
   {}

   \maketitle
   
%

\section{Introduction}
With the development of high-contrast and high-resolution direct imaging, a variety of structured protoplanetary disks have been revealed \citep{Benisty2023}. Multiwavelength polarimetric differential imaging (PDI) offers a unique perspective from which to characterize the dust scattering properties. Among the polarization observables, the degree of polarization of the scattered light is particularly valuable because it is less sensitive to the overall disk structure, thus allowing for more accurate diagnostics of dust properties, such as grain size and porosity\citep{Min2016, Tazaki2022}, which directly influences planet formation processes \citep{Garcia2020}. 
Studying dust evolution is critical because grain growth depends on local disk conditions, such as turbulence, temperature, and interactions with companions and embedded planets \citep{Birnstiel2024}. 

HD~100453 is an ideal system to investigate to gain insight into environment-dependent dust evolution. HD~100453A is a Herbig A9Ve star in the lower Centaurus Association, located at 103.8 pc\citep{Gaia2023}, with a close M-type companion, HD~100453B, separated by 1.05'' \citep{Collins2009}. The main star shows an IR-excess typical for a transition disk with strong emission from hot dust around 2-5~$\mu$m, another maximum around 25~$\mu$m from cold dust,
and polycyclic aromatic hydrocarbons emission bands \citep{Malfait1998, Meeus2001}. The cold dust component
is well resolved in scattered light, showing a disk with a cavity of $r=0.14$\arcsec, a bright rim inclined by 38$^\circ$, two symmetric trailing spiral arms, and two shadows on opposite sides of the disk ring in the NE and SW \citep{Wagner2015, Benisty2017}. The two shadows are caused by a misaligned inner disk, which accounts for the hot dust emission in the near-IR.
Atacama Large Millimeter/submillimeter Array (ALMA) Band 6 revealed CO isotopolog emissions, showing gas presence within the disk cavity \citep{vanderplas2019} and proved that the disk is gas poor \citep{vanderplas2019, Collins2009}. The subsequent Band 7 reached a higher angular resolution and detected spirals in both CO and dust emission, with the southern spiral connecting to the companion \citep{Rosotti2020}.
These features make HD~100453 a well-observed and studied system across a wide range of wavelengths. 

The origins of the spirals and shadows in HD~100453 have been widely investigated. 
Popular explanations include the massive companion, the optically thick misaligned inner disk, and undetected planets. 
Previous works have characterized the system's spiral morphology and companion properties \citep{Benisty2017, Wagner2018, Gonzalez2020}, yet a detailed analysis of dust properties in the disk, particularly across its substructures comparing the ring, spirals, and cavity, is lacking. Understanding the dust characteristics is critical, as different dynamical environments within the disk can significantly alter the grain size distribution through processes such as dust filtration \citep{Zhu2012, Pinilla2016} and spiral-induced dust concentration \citep{Dipierro2015, Cuello2019}. 
We aim to use multiwavelength polarimetric observations to reveal variations in dust grains across different disk regions, providing insight into how disk-companion interaction influences dust growth.

In this paper, we present multiwavelength polarimetric imaging, including the first $V$-band observation and new observational data at the $I'$, $J$, and $Ks$ bands for HD~100453, and combine them with the archival $Ks$ band in order to investigate the dust properties and explore the variability seen in scattered light. In section~\ref{sect: observation} we describe the selected observations and the data reduction procedure. In section~\ref{sect: analysis}, we measure the disk-integrated polarized flux, the disk intensity, and the fractional polarization considering the point spread function (PSF) smearing effect. In section~\ref{sect: model}, we use the radiative transfer (RT) RADMC-3D disk model to constrain the dust scattering properties in different regions. In section~\ref{sect: discussion}, we discuss dust size, structure, and composition in the disk and relate them to the forming environment. We summarize the main results in section~\ref{sect: conclusion}. 

\begin{table*}[]
    \centering
    \caption{VLT/SPHERE polarimetric imaging data and corresponding observing conditions.}
    \begin{tabular}{l l c c c c c c c}
    \hline
    \hline
    Epochs          & Filters         &  Catg. & nDIT $\times$ DIT & $n_{cyc}$  &  seeing[$\arcsec$] & $\tau_0$[ms] & FWHM[mas] & Label \\
    \hline
    Feb 02, 2023    & $V$      & Object           & 10 $\times$ 5 s      & 8     & 0.51-0.77 &  7.8-11.0 & 26.3 & V\\
                    & $I'$     & Object           & 12 $\times$ 2 s      & 8     & 0.54-0.72 &  6.2-9.1  & 26.5 & I\\ 
    Mar 02, 2023    & $J$ + ND\_1.0      & Object  & 20 $\times$ 0.8375 s   & 8   & 0.44-1.26 & 4.5-9.8 & 47.1 & J\\
                    & $Ks$ + ND\_1.0     & Object      & 20 $\times$ 0.8375 s   & 8   & 0.48-0.95 & 5.4-11.7 & 63.4 & K\smallskip \\
    Mar 03, 2023    & $V$      & Object           & 6 $\times$ 10 s      & 9     & 0.54-1.20  &  3.4-8.7   & & V-hop\\  
                    &  $V$         & Object           & 2 $\times$ 5 s       & 3     & 0.81-0.94 &  4.5-5.7 & 24.5\\
                    &  $V$        & Reference \tablefootmark{a}       & 6 $\times$ 10 s      & 3     & 0.81-1.02 &  3.7-5.9 \\ 
    Jun 09, 2022\tablefootmark{c}    & $Ks$ + N\_ALC\_Ks    & Object                        & 1 $\times$ 32s         & 23\tablefootmark{b}  & 0.60-1.09 & 5.3-11.3 &   & K-hop\\
                    &  $Ks$ + ND\_2.0  & Flux        & 5 $\times$ 2s          & 2   & 0.70-0.74 & 6.4-8.0 & 68.8 \\
                    & $Ks$ + N\_ALC\_Ks & Reference \tablefootmark{a}  & 1 $\times$ 32s         & 6   & 0.66-1.07 & 4.2-9.2\\
    \hline
    \end{tabular}
    \tablefoot{
        {Here, nDIT $\times$ DIT gives the exposure time for one out of four half-wave plates settings for a polarimetric cycle. Thus the total integration time for one polarimetric instrument setting is $nDIT \times DIT \times 4 \times n_{cyc}$}.
        \tablefoottext{a}{The reference stars are HD101582 for V-hop and HD100541 for K-hop sequences.} \tablefoottext{b}{Two cycles are discarded.} \tablefoottext{c}{Archival and published in \citet{Ren2023}}.} 
    \label{tab:obs-info}
\end{table*}

\section{Observations and data reduction}
\label{sect: observation}
\begin{figure*}
    \centering
    \includegraphics[width=\textwidth]{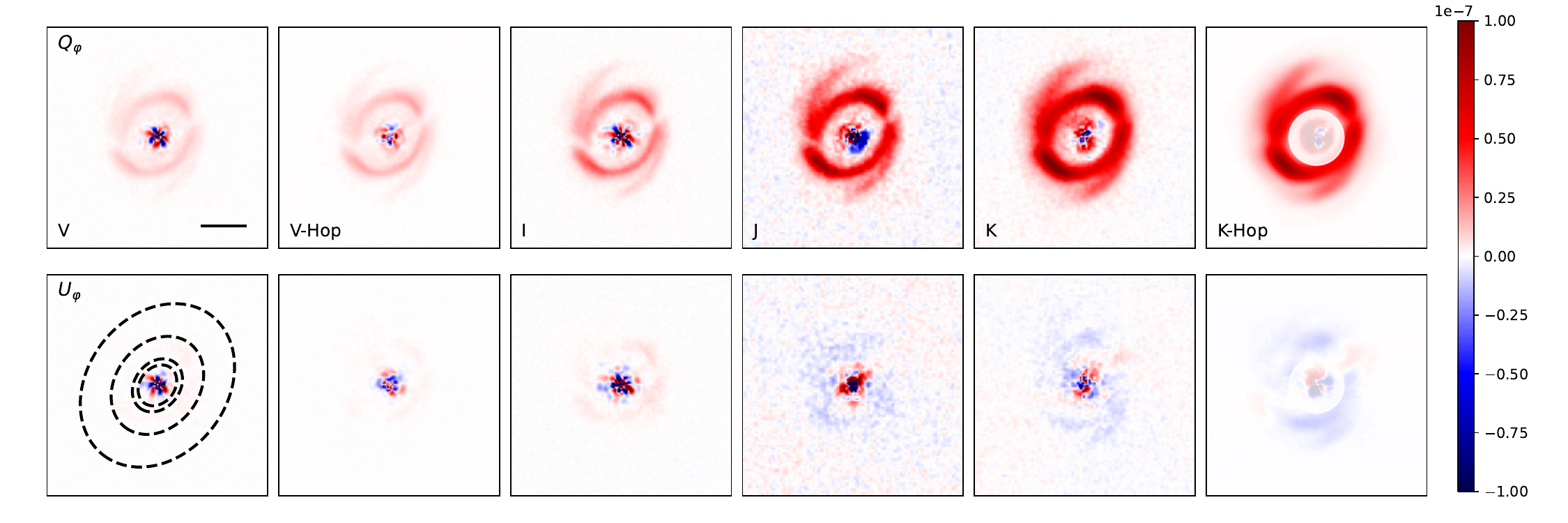}
    \caption{Polarized intensity images $Q_{\varphi}(x,y)$ and $U_{\varphi}(x,y)$ for $V$, $I'$, $J$, and $Ks$ band. All images are normalized to the integrated stellar intensity $I_{\star}$ and their pixel scale ($3.6\times3.6$ mas$^2$ for ZIMPOL and $12.25\times 12.25$ mas$^2$ for IRDIS). The shared color bar is indicated on the right. All the images are cut to $1.0\times1.0\arcsec$ and the black bar in $Q_{\varphi}(V)$ indicates 0.2\arcsec. The dashed lines in $U_{\varphi}(V)$ enclose the integration region for the cavity, the rim, and the spiral respectively. The semitransparent circle in K-hop represents the inner working angle (IWA) = 0.125\arcsec limited by the coronagraph. North is
    up and east to the left.}
    \label{fig: observations}
\end{figure*}

\subsection{Selected observations}
New HD 100453 data were taken in 2023 (ESO Program ID: 110.245U.001, PI: J.~Ma) using the SPHERE/VLT high contrast adaptive optics instrument \citep{Beuzit2019}. This observation is specifically 
designed for accurate multiwavelength, photo-polarimetric measurements of the scattered light from the disk to characterize dust properties.
The data include nonsaturated, short integrations in the $V$, $I'$, $J$, and $Ks$-bands, and longer integrations with saturation at $V$ using a star-hopping procedure to calibrate the PSF with reference star observations.
All the observations are taken in P1 mode with field rotation to correct for fixed instrument features, reach a higher precision for polarimetry, and facilitate angular differential imaging for the disk intensity extraction.

Nonsaturated observations were carried out for all four wavelengths, using fast read-out mode and short exposure time for the $V$ and $I'$-band observations with ZIMPOL, and neutral density (ND) filters for the $J$ and $Ks$-bands observations with IRDIS. Such nonsaturated observations provide simultaneous measurements of the polarized intensity of the disk and the flux in the PSF peak of the central star to monitor and account for temporal variation of the PSF convolution in the data \citep{Tschudi2021}.

The star-hopping sequence at $V$ (labeled V-hop) is designed to switch between long-exposures and short-exposures on target, combined with hopping to the PSF reference star for long-exposures. 
We designed three iterations of hopping, each containing three polarimetric cycles of long exposures and one polarimetric cycle of short exposures on HD~100453, and one polarimetric cycle of long exposure on reference star. The long exposures on science target and reference stars are saturated in the center, but deep enough to allow disk intensity extraction. Short exposures in between monitor the stellar (and disk) variability and the seeing condition. 
Details of the observing conditions and PSF quality parameters are summarized in Table~\ref{tab:obs-info}. 
Analyzing reflected light from compact disks regions, as in this case, is very difficult with high precision polarimetry, as even small PSF variations can significantly affect the measured polarization signal \citep{Tschudi2021}. Therefore, each epoch of observation is designed to have (quasi-) simultaneous unsaturated PSFs so that fast PSF variations can be taken accurately into account. 

This study also includes archival SPHERE $Ks$-band data of HD~100453 taken in June 2022 from a coronagraphic $Ks$-band survey program on protoplanetary disks using star hopping (Program ID: 105.20JB.001, PI: M.~Benisty). 
The hopping sequence for HD~100453 lasted for more than two hours, starting and ending with the science target, and hopping to a reference star for six times. For each hop on the science target,
two or three long polarimetric cycles were taken with coronagraph N\_ALC\_Ks. The observation does not include short exposures in between, only two flux frames were taken with ND filter, but without polarimetric optical components, one before and one after all the polarimetric cycles. 
The observational results of this program are published in \citet{Ren2023}.

\subsection{Data reduction}
\label{sect:reduction}
ZIMPOL data were reduced with the IDL pipeline sz-software as described in \citet{Schmid2018}, and the IRDIS data were reduced with IRDAP (IRDIS Data reduction for Accurate Polarimetry) pipeline \citep{vanHolstein2020}. For each raw data frame, these dedicated pipelines perform basic image extraction, bad pixel correction, dark subtraction, flat-fielding, and centering steps. The sz-software allow for an extra beamshift correction, which is required for high-performance polarimetric data. The pipelines perform standard double-difference to derive the Stokes parameters $Q = I_{0}-I_{90}$, $U=I_{45}- I_{135}$, and double-addition methods to compute total intensities $I_{Q} = I_{0} + I_{90}$, $I_{U} = I_{45} + I_{135}$. We define the final total intensity as the average $I = (I_Q+I_U)/2$ . 

On top of the data reduction with their dedicated pipelines, we apply an extra step to subtract the stellar polarization. 
For all integrations without saturation,
we measure the fractional polarization $p = \sqrt{(Q/I)^2 + (U/I)^2}$ and position angle $\theta = 1/2\cdot\arctan(U/Q)$ in a $r=0.05\arcsec$ round aperture that encloses the stellar PSF core. For long-integrations in $V$-band when the center is saturated, the stellar polarization is measured in an annulus covering the radius range [0.29, 0.36]$\arcsec$ of the speckle ring at about $20\lambda/D$ which is mostly caused by light from the central source, but also contains small contributions from the polarization signal from the disk.

For $Ks$-band long-integration, the stellar polarization is measured between [0.98, 1.49]$\arcsec$ in the speckle ring at this wavelength, which is located outside the disk emission, and the SE quadrant is excluded from the integration region to avoid contamination from the companion. We note that the unresolved central source contains the signal from the inner disk.

We measured for HD~100453A a very low degree of polarization $p<0.2\%$ at $V$, $I$, and $J$, indicating that contributions
from intrinsic stellar polarization, instrumental polarization, and interstellar polarization are small. A low residual instrumental polarization is typical for SPHERE polarimetry, and also a
low interstellar polarization is expected for
HD~100453 because of its low interstellar 
extinction \citep{Malfait1998}.
The $Ks$-band shows a higher polarization $p=0.6\%$ with a position angle
of about $\theta\approx 170^\circ$.
This is compatible with an intrinsic polarization perpendicular to the orientation of the major axis of the unresolved hot disk of $80^\circ$ according to the model of \citet{Benisty2017}, which dominates the central source emission in the $Ks$-band.
Unfortunately, the uncertainties in this interesting polarization result are not well defined and would require much improved instrument calibration procedures to correct residual instrumental effects and achieve reliable values at the $\Delta p<\pm 0.1~\%$ level.
Therefore, we subtracted the measured stellar polarization $(p, \theta)$ at each wavelength from the Stokes $Q$ and $U$ maps using $Q(x,y)=Q_{\rm obs}(x,y)-I(x,y)p\cos \theta$ and $U(x,y)=U_{\rm obs}(x,y)-I(x,y) p\sin \theta$ so that the central star has zero polarization \citep{Schmid2025}.

For $Ks$-band observations, we carried out an extra background subtraction step on the intensity frames, which is necessary for later quantitative analysis. And for K-hop observation, we applied a transmission correction for the coronagraph N\_ALC\_Ks, which is indispensable for compact disks such as HD~100453 whose inner edge is close to the inner working angle(IWA) = 0.125\arcsec of the coronagraph. These steps are detailed in the Appendix~\ref{app: bck-subtr} and \ref{app: coronagraph-correction}.

As the last step, we calculated the azimuthal polarization parameters $Q_{\varphi}(x,y)$ and $U_{\varphi}(x,y)$ maps from the reduced Stokes polarization $Q(x,y)$ and $U(x,y)$ maps:
\begin{align}
    Q_{\varphi}(x,y) & = -Q(x,y)\cos(2\varphi) - U(x,y)\sin(2\varphi) , \\
    U_{\varphi}(x,y) & = Q(x,y)\sin(2\varphi) - U(x,y)\cos(2\varphi) ,
\end{align}
where $\varphi$ is the position angle of the map location $(x,y)$ to the central star from north over east. Fig.~\ref{fig: observations} presents the mean $Q_{\varphi}(x,y)$ and $U_{\varphi}(x,y)$ maps for the different data sets. 

\section{Data analysis}
\label{sect: analysis}
\begin{table*}[]
    \caption{Observed polarized flux $Q_{\varphi}/I_{\star}$ integrated in different apertures for HD~100453 for each observing date.}
    \label{tab: polarised-flux}
    \centering
    \begin{tabular}{c c c c c c c c}
    \hline
    \hline
    Label    & $Q_{\varphi}/I_{\star}$ & $\hat{Q}_{\varphi}/I_{\star}$ & $\hat{Q}^{\rm rim}_{\varphi}/I_{\star}$ & $\hat{Q}^{\rm sp}_{\varphi}/I_{\star}$ & $Q^{\rm cav}_{\varphi}/I_{\star}$ \\  
    & \% & \% & \% & \% & \% \\
    \hline
    V    & 0.099 $\pm$ 0.006  & 0.291 $\pm$ 0.005  & 0.188 $\pm$ 0.002 & 0.056 $\pm$ 0.002 &  0.005 $\pm$ 0.001\\ 
    I    & 0.230 $\pm$ 0.006  & 0.444 $\pm$ 0.004  & 0.300 $\pm$ 0.003 & 0.109 $\pm$ 0.002 & 0.006 $\pm$ 0.001\\ 
    J    & 0.611 $\pm$ 0.056  & 0.891 $\pm$ 0.077 & 0.631 $\pm$ 0.011 & 0.278 $\pm$ 0.018 & 0.013 $\pm$ 0.002\\ 
    K    & 0.958 $\pm$ 0.046  &  1.212 $\pm$ 0.052 & 0.742 $\pm$ 0.005 & 0.388 $\pm$ 0.005 & 0.017 $\pm$ 0.004 \smallskip \\ 
    V-hop\tablefootmark{a}     & 0.094 $\pm$ 0.008 & 0.300 $\pm$ 0.015 & 0.189 $\pm$ 0.002 & 0.055 $\pm$ 0.002 & 0.006 $\pm$ 0.001 \\ 
    V-hop\tablefootmark{b}    & 0.125 $\pm$ 0.022  & 0.35 $\pm$ 0.06 & - & - & - \\ 
    K-hop\tablefootmark{c}    & 0.871 $\pm$ 0.054 & 1.25 $\pm$ 0.08 & - & - & - \\ 
    \hline
    \end{tabular}
    \tablefoot{From left to right, each column represents the polarized flux integrated in the entire disk region; the entire disk region, corrected for convolution effect; the rim region, corrected for convolution effect; the spiral region, corrected for convolution effect; and the cavity region. 
    The disk, rim, spiral, and cavity regions are defined as elliptical annulus from [0.10, 0.86]\arcsec, [0.13, 0.24]\arcsec,[0.24, 0.40]\arcsec, and [0.10, 0.13]\arcsec, respectively. 
    \tablefoottext{a}{Short integrations in the star-hopping sequence.}\tablefoottext{b}{The last cycle short exposure total intensity in V-hop is used as the representative PSF for smearing correction.} \tablefoottext{c}{The central 125mas is masked out.}}
\end{table*}

\subsection{Stellar intensity}
HD~100453A is classified as a Herbig Ae/Be star, which are young pre-main-sequence stars known for variability. Photometric
results from the Hipparcos satellite have shown
variations with a scatter in the
star brightness of $\pm 0.026$~mag during about
3.5 years. We neglect such variations 
for our photo-polarimetric study,
which is mostly based on relative measurements
between polarized flux signals and the intensity of the system HD~100453A.

We used the total intensity of the main star HD~100453A as the reference value $I_\star$, including the contributions from the circumstellar disk but with the companion filtered out. 
For this we first calculated the azimuthally averaged radial profile $I_{\star}(r)$ and its standard deviation $\Delta I_{\star}(r)$ in the annulus section, defined by $r=[0.6\arcsec, 1.6\arcsec]$ 
and $\varphi=[0, 90^\circ]\bigcup [180^\circ, 360^\circ]$ excluding the quadrant regions with
the companion. We then replace in the quadrant
of the companion $\varphi=[90^\circ, 180^\circ]$ all points $I(r,\varphi)$
with intensities outside the range $I_{\star}(r)\pm\Delta I_{\star}(r)$, typically with a higher value because of the additional contribution from the companion,
by a random number drawn from a Gaussian distribution with mean $I_{\star}(r)$ and standard deviation $\Delta I_{\star}(r)$.

The intensity $I_\star$ for HD~100453A is then integrated within the round aperture of 3$\arcsec$-diameter for all nonsaturated images. 
The uncertainty for $I_\star$ is the standard deviation for all intensity frames for a given instrument setting. For K-hop, only two flux frames are available and we use half of the difference between the two as uncertainty for $I_\star$. The resulting uncertainty is $2\%$, consistent with deviations obtained for other settings.

We used the companion-subtracted images for HD~100453A as a proxy for the PSF profiles of individual observations, which are used to correct PSF smearing effects. 
The contribution of the extended disk intensity to the $I_\star$ value is between 1~\% and 4~\%, as is derived later. 
Temporal variations in atmospheric turbulence change the shape of the PSF profile, much more than the flux contribution of the disk. Therefore, using the measured intensity distribution for HD 100453 as PSF for the convolution correction will have only a small effect on the derivation of the intrinsic polarization signal. 

In addition, the difference between the total intensity frame and the filtered intensity frame gives the intensity of the companion. We derive
the flux ratio between the B and A components about 0.05\% at $V$ and 1.25\% at $Ks$. More details for
all used filters and magnitudes are given in appendix~\ref{app: companion}. 

\subsection{Polarized disk intensity}
\begin{figure}
    \centering
    \includegraphics[width=0.48\textwidth]{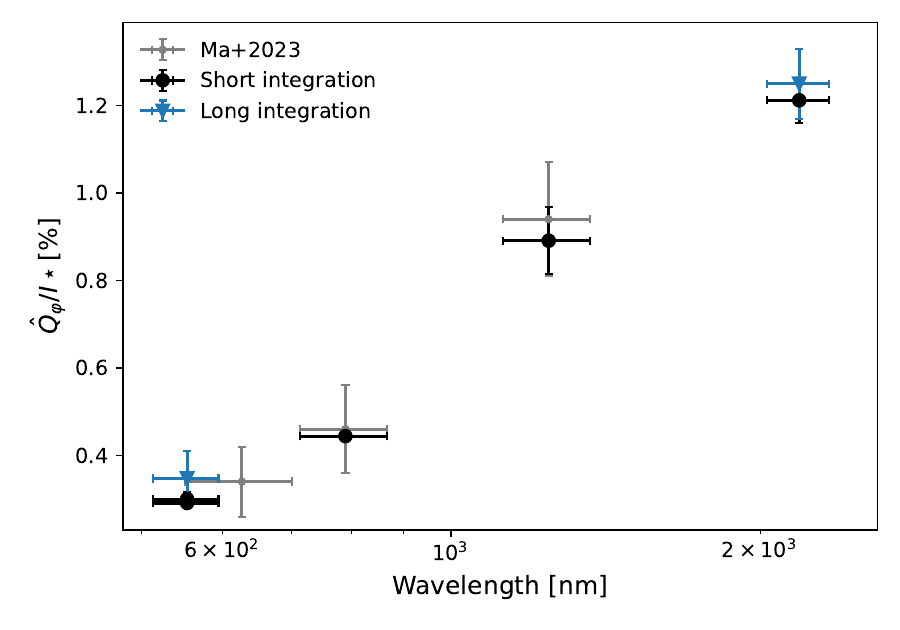}
    \caption{Intrinsic integrated polarized flux $\hat{Q}_{\varphi}/I_{\star}$ as a function of wavelengths. Values are from Table~\ref{tab: polarised-flux}. }
    \label{fig: reflectivity}
\end{figure}

The measurement of the azimuthal polarization $Q_{\varphi}$ of the circumstellar scattering regions depends on the spatial resolution of the observations. The main effect of PSF smearing in polarimetry is the cancellation between positive and negative signals in the Stokes $Q$ and $U$ quadrants caused by scattering. Therefore, one needs to take the PSF convolution into account to derive the intrinsic photo-polarimetric signal.
All short exposures obtained in 2023 provide the polarization signal and the unsaturated intensity signal simultaneously. This allowed us to correct for the variable PSF smearing effects cycle by cycle following the procedure described in \cite{Ma2024}. 

The correction relies on the fact that convolution is a linear operation. The PSF smearing of an inclined disk can be simulated by summing the convolutions of narrow concentric ellipses that align with the inclined geometry. For a given PSF, the smearing is stronger at small radii and weaker at larger radii. Using this principle, we derive a two-dimensional correction map for each polarimetric cycle to correct the observed disk images.
Simulations in \cite{Ma2024} show that this approach recovers the PSF-smearing-corrected intrinsic polarization signal $\hat{Q}_\phi/I_\star$ with an error of less than a few percent for disks with roughly an ellipse-like appearance such as HD~100453.

For our analysis, we derive the intrinsic azimuthal polarization  $\hat{Q}_\phi/I_\star$ within the elliptical aperture, with radius range of [0.10, 0.86]\arcsec inclined by $38^\circ$, as shown in Fig.~\ref{fig:mask-r-k} for each polarimetric cycle of all instrument settings. The results are listed
in Table~\ref{tab: polarised-flux} and plotted in Fig.~\ref{fig: reflectivity}. 

For the V, I, J, K observations and V-hop short integrations with simultaneous $Q_{\varphi}$ and $I_{\star}$ measurements, we report the mean $\hat{Q}_{\varphi}/I_{\star}$ value and standard deviation of all cycles.
For the K-hop data where $I_{\star}$ and $Q_{\varphi}$ were obtained separately, error propagation is applied to derive the final $Q_{\varphi}/I_{\star}$ and $\hat{Q}_{\varphi}/I_{\star}$ values.
We note that for K-hop, the central 125~mas region is masked out, so the resulting values represent lower-limit estimates.
The V-hop long-integrations lack simultaneous unsaturated PSF information and the seeing conditions varied over the two-hour observation. This explains the large standard deviation $\Delta Q_{\varphi}/Q_{\varphi} = 16\%$. 
Fortunately, nonsaturated short integrations taken in between the long integrations helped to account for the seeing variations, though the relative uncertainties remain larger than those in the other datasets.
Fig.~\ref{fig: reflectivity} includes the measurement in the $R'$, $I'$, and $J$-bands from \cite{Ma2024} using the same integration aperture. The excellent agreement between the different measurements indicates that the azimuthal polarization signal of HD~100453 is stable.

We also measured the integrated flux from disk substructures using concentric elliptical apertures: the cavity, the rim, and the spirals, corresponding to the radial ranges of [0.10, 0.13]\arcsec, [0.13, 0.24]\arcsec, and [0.24, 0.40]\arcsec respectively for epochs that have simultaneous PSFs. Flux correction was applied to observed polarization for the rim and spirals regions, and the resulting intrinsic values $\hat{Q}^{\rm rim}_{\varphi}/I_{\star}$ and $\hat{Q}^{\rm sp}_{\varphi}/I_{\star}$ are summarized in Table.~\ref{tab: polarised-flux}. The $\hat{Q}^{\rm sp}_{\varphi}/I_{\star}$ shows a faster increase with wavelength than the $\hat{Q}^{\rm rim}_{\varphi}/I_{\star}$, hinting at different dust species in the substructures, which we discuss in Section~\ref{sect: dust-properties}. 
We list only the observed integrated polarized flux from the cavity $Q^{\rm cav}_{\varphi}/I_{\star}$ in Table.~\ref{tab: polarised-flux} because the innermost region is too faint and too close to the star for a reliable PSF-smearing correction using the method in \citet{Ma2024}. Instead, we correct the cavity flux using the forward-modeling approach described in Section~\ref{sect: set-up-c}. We note that the substructure apertures do not fully cover the total flux integration region, which was chosen conservatively to include all smeared signal, so their sum does not equal the total flux. 

\subsection{Disk intensity}
The V-hop and K-hop observing sequences used a star-hopping strategy \citep{Wahhaj2021}, taking quasi-simultaneous total intensity observations of a PSF reference star
enabling reference differential imaging to extract the disk intensity map for HD 100453 A. 
The reference star is chosen to have a similar spectral type and magnitude as the science target but hosts no disk. 
For the first time, this study applies this method to observations with the SPHERE/ZIMPOL camera.
For V-hop, we obtained 54 total intensity images on target $I_{\rm tar}(x,y)$ and 18 images of total intensity images on reference star $I_{\rm ref}(x,y)$. For K-hop, we treated $I_Q$ and $I_U$ separately to match the number of image with $V$-band, resulting in 46 target images and 12 reference images in total. 

The disk signal can be extracted from the difference
\begin{equation}
\label{eq: disk-intensity}
    I_{\rm disk}(x,y) = I_{\rm tar}(x,y) - a\cdot I_{\rm ref} (x,y) ,
\end{equation}
where $a$ is a constant scaling factor optimized to minimize the residuals within the defined fitting region with preferentially no contributions from the disk for each reference image. 
The residual is defined as the sum of the relative squared differences in each pixel over the fitting region $\Sigma$:
\begin{equation}
    R = \sum_{x,y} (I_{\rm tar}(x,y) - I_{\rm fit}(x,y))^2/I_{\rm tar}(x,y) ,
\end{equation}
where the fit image, $I_{\rm fit}$, is defined as
\begin{equation}
    I_{\rm fit}(x,y) = a\cdot I_{\rm ref} (x,y) .
\end{equation}
Ideally, the fitting region, $\Sigma$, should exclude any disk signal. The speckle ring is usually a good choice for this fitting region because it contains a strong stellar signal while minimizing the contributions from the disk.
However, the PSF in the $V$-band introduces a nonnegligible extended disk halo component, $I_{\rm halo}$, that spreads the disk signal into the surrounding area (Fig.~\ref{fig:model-v-k}). This smeared component must be considered to avoid oversubtraction of disk intensity for the $V$-band. A quantitative analysis of this effect is provided in appendix~\ref{app:disk-intensity}, where we demonstrate that even a small contamination from the halo in the fitting region can cause significant errors when the disk is faint compared to the stellar PSF. Therefore, for $V$-band, we modified the fitting function to include the halo component:
\begin{equation}
    I_{\rm fit} (x,y) = a\cdot I_{\rm ref} (x,y) + I_{\rm halo}(x,y) .
\end{equation}
The fitting regions are shown in Fig.~\ref{fig:mask-r-k}. For V-hop, we adopted an annulus between [0.43, 0.76]\arcsec, chosen to include sufficient stellar signal while avoiding bright spirals. The halo component $I_{\rm halo}(x,y)=I_{\rm halo}(r)$ is assumed to be radially symmetric, and is fitted using the disk model Setup A (introduced in Section~\ref{sect: model}), which follows exponential fall-off with radius. 

For K-hop, we estimated that the halo component of the disk intensity is negligible, and we set $I_{\rm halo} = 0$ in the fitting process. The fitting region is defined as an annulus between [0.98, 2.14]\arcsec, with two circular apertures excluded to mask the companion and the known background star in the NE direction \citep{Collins2009, Wagner2015}. 

\begin{figure}
    \centering
    \includegraphics[width=0.98\linewidth]{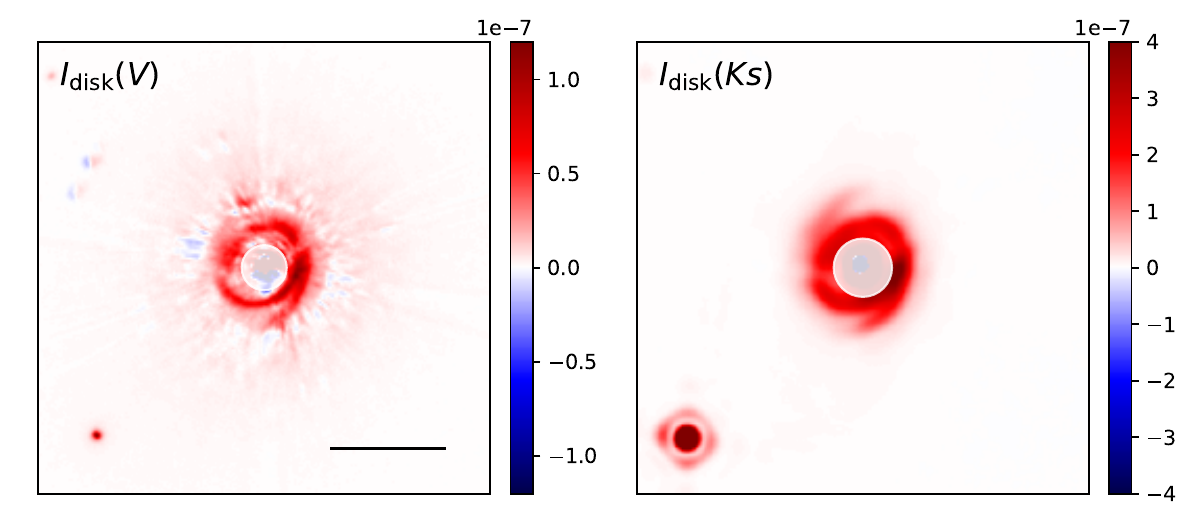}
    \caption{Extracted disk intensity map $I_{\rm disk}(x,y)$ for the $V$- and $Ks$-bands, cut to $2.0\times 2.0\arcsec$. The black bar indicates 0.5\arcsec. The central 0.1\arcsec and 0.125\arcsec are masked out as the IWA for $V$ and $Ks$ respectively. }
    \label{fig:disk-intensity-vk}
\end{figure}

For a given target image $I_{\rm tar}$, we search for the best-fitting scaling value $a$ for each reference intensity image $I_{\rm ref}$ and evaluate the result using the residual. We average the three best-fitting results among all scaled-$I_{\rm ref}$ and obtain $I_{\rm best}(x,y)$-image for this given target image $I_{\rm tar}$. We apply this process for each target image and obtain 54 and 46 $I_{\rm best}(x,y)$-image for $V$ and $Ks$-band, respectively.
Among all $I_{\rm best}$ images at given wavelength, we select the 10 with the lowest residuals (representing the top 20\% in PSF subtraction quality) and compute their average to produce the final disk intensity map $I_{\rm disk}(x,y)$, shown in Fig.~\ref{fig:disk-intensity-vk}. The standard deviation among the selected images is used as a conservative uncertainty estimate, which is primarily driven by the seeing variability. 

The extracted total disk intensity resembles the structure seen in polarized intensity. At $V$-band, the disk signal partly coincides with the strong
speckle ring from the PSF near 0.4\arcsec, and therefore, the extracted intensity signal is noisy. The $Ks$-band disk intensity is less affected by speckle noise. In both bands, the companion and the background star are recovered. 

For the integrated intensity, we used the same elliptical integration region as for the polarized intensity. We integrated the disk total intensity in the best 10 $I_{\rm disk}(x,y)$ images, with the mean and standard deviation taken as the final flux and uncertainty. 
The measured disk intensity values are $I_{\rm disk}/I_\star= 1.2\pm 0.3$~\% for V-hop and $3.8\pm 0.5$~\% for K-hop.  
We note that for K-hop data, we adopt IWA=0.125\arcsec, so the integrated flux $I_{\rm disk}$ represents a lower limit. Moreover, the
$I_{\rm disk}$ values are not yet corrected for the PSF convolution, which redistributes substantial disk intensity signal to regions outside the measuring aperture. This effect is quantified with convolved disk models in Section~\ref{sect: model} to derive intrinsic disk parameters.

\subsection{Degree of polarization}
\begin{figure}
    \centering
    \vspace{-0.2cm}
    \includegraphics[width=0.98\linewidth]{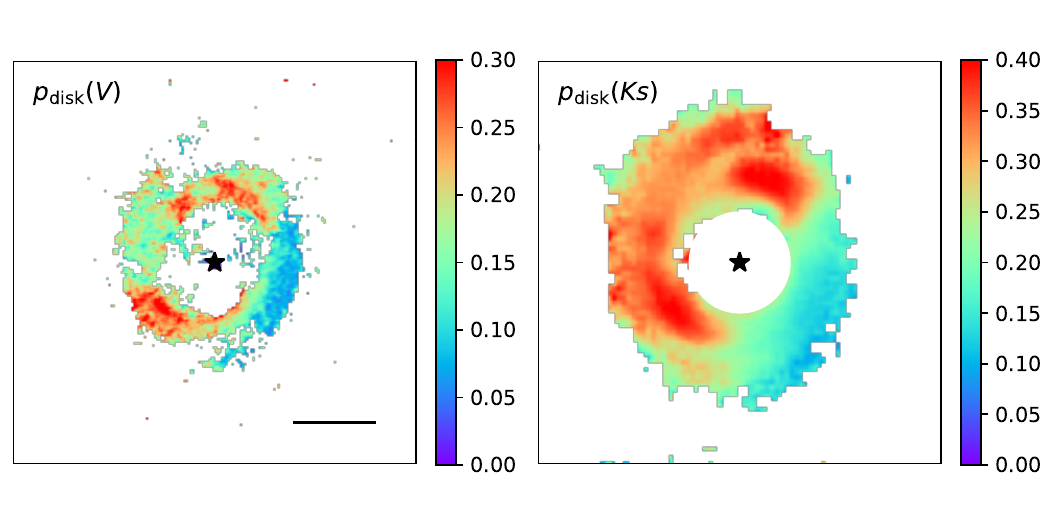}
    \vspace{-0.3cm}
    \caption{Observed degree of polarization $p_{\rm disk}(x,y)$ for $V$- and $Ks$-bands cut to $1.0\times 1.0\arcsec$. The black bar in the $V$-band subfigure indicates 0.2\arcsec. The star marker indicates the position of the central host star. }
    \label{fig: pdisk-v-k}
\end{figure}

From the observed polarized intensity and total intensity maps for the disk, we derive the degree of polarization map as $p_{\rm disk}(x,y) = Q_{\varphi}(x,y)/I_{\rm disk}(x,y)$. For each of the 10 best-fitting maps $I_{\rm disk}(x,y)$, we use the corresponding cycle of $Q_{\varphi}(x,y)$ to calculate $p_{\rm disk}(x,y)$. The final result is the mean of these 10 maps, shown in Fig.~\ref{fig: pdisk-v-k}, zoomed into $1 \times 1\arcsec$. To highlight the substructures, different color scales are applied for two bands. The standard deviation among the 10 images is taken as the uncertainty $\sigma(p_{\rm disk})$. 
Pixels with low disk signal and high uncertainties exceeding $100\%$ (for $V$-band) and $50\%$ (for $Ks$-band) of the mean value are not shown. These maps present the convolution-smoothed spatial distribution of $p_{\rm disk}$. While the regions with strong disk signals are less affected by systematic errors, the uncertainties per pixel remain large with deviations $\pm \Delta p_{\rm disk}/p_{\rm disk}$ of 40\% for $V$-band and $20\%$ for $Ks$-band.

The key observable derived from the $p_{\rm disk}$ maps is the maximum degree of polarization ${\rm max}(p_{\rm disk})$. For a disk with a given surface structure, this ${\rm max}(p_{\rm disk})$ typically occurs where the scattering angle is close to $90^\circ$, producing a high degree of polarization \citep{Ma2022}.
Such favorable scattering angles naturally arise in most disk geometries, making ${\rm max}(p_{\rm disk})$ a robust and widely applicable diagnostic for characterizing dust scattering properties across different wavelengths and disk types \citep{Tazaki2022}.

For the determination of ${\rm max}(p_{\rm disk})$, we select the top 10~\% of pixels with uncertainties $\pm \Delta p_{\rm disk}/p_{\rm disk}$ below $20\%$ of the mean value. The selected pixels are concentrated on the SE side of the major axis at $V$-band, and distributed on the major axis and the spiral tail at $Ks$-band. 
The final ${\rm max}(p_{\rm disk})$ values are taken as the average over these pixels, with the uncertainty dominated by the pixel-wise variation. The measured ${\rm max}(p_{\rm disk})$ are $28\pm 2$\% and $39 \pm 4$\% respectively. 

We note that the $p_{\rm disk}$ represents observed values, as both $Q_{\varphi}(x,y)$ and $I_{\rm disk}(x,y)$ are smeared by PSF convolution. 
Despite this limitation, several interesting features are already visible in the observed $p_{\rm disk}$ maps. 
In the $p_{\rm disk}(V)$ map, the SE side shows a higher degree of polarization ($28\%$) compared to the NW side ($21\%$). The origin of this asymmetry remains unclear. 
In the $p_{\rm disk}(K)$ map, the NE spiral exhibits an enhanced degree of polarization.  
In both $p_{\rm disk}(V)$ and $p_{\rm disk}(K)$ maps, the shadowed region exhibits a slightly lower degree of polarization, leading to discontinuities in the azimuthal variance along the rim. 
It is also important to note that, although the coronagraph leads to signal loss in both polarized intensity and total intensity, the $p_{\rm disk}$ is not affected at all as long as the transmission function is rotationally symmetric, because the same transmission correction map applies to $Q_{\varphi}$ and $I_{\rm disk}$. 

We also measured the disk-averaged degree of polarization $\langle p_{\rm disk}\rangle = 16\pm 5\%$ at $V$-band and $33\pm6\%$ at $Ks$-band. These values are derived directly from the observed polarization and intensity $\hat{Q}_{\varphi}/I_\star$ in the given elliptical aperture and does not consider that the PSF convolution effects are different for the $Q_\phi$ and $I$ disk signal \citep{Schmid2025}. This problem is taken correctly into account with the forward modeling results from Sect.~\ref{sect: model} providing intrinsic disk values.

\subsection{Radial profile and brightness ratio}
\begin{figure}
    \centering
    \includegraphics[width=0.48\textwidth]{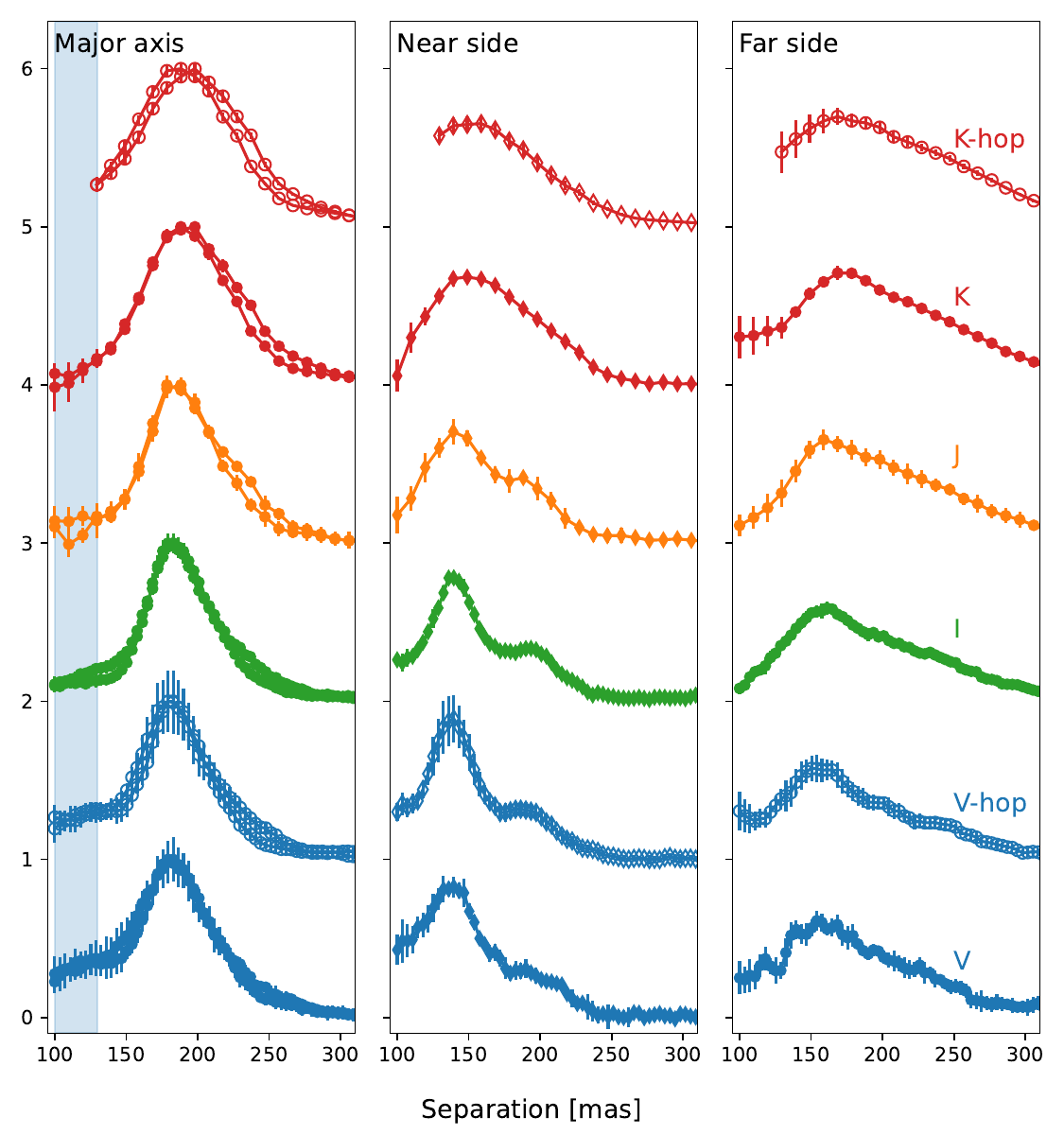}
    \caption{Radial profiles of polarized intensity $Q_{\varphi}(r)$ along both sides of the major axis (left), near side (middle), and far side (right). All profiles have been normalized to the peak value along the major axis and shifted by one for clarity. The standard deviation among all cycles at a given epoch is indicated as the error bar. The shaded region indicates the cavity region [0.10, 0.13]\arcsec. }
    \label{fig:radial_profiles}
\end{figure}

To characterize the scattering behavior of dust grains in the disk in more detail, we calculate the radial profiles within $30^\circ$ wedges centered at position angle $= 142^\circ$ and $322^\circ$ (major axis), $232^\circ$ (near side), and $52^\circ$ (far side). The mean profiles for each instrument setting are plotted in Fig.~\ref{fig:radial_profiles}. 
$V$ and $Ks$ bands have two epochs of observations. 
The error bars represent the standard deviations between 
the profiles for all cycles of a given instrument setting. 
All profiles are normalized to the peak value along its major axis. 

To quantify the observed asymmetries, we define brightness ratios based on these profiles (Table~\ref{tab:contrast}).
The near-to-far side ratio $R_{N/F}$ is defined as the peak brightness on the near side divided by that on the far side. The near-to-major axis ratio $R_{N/M}$ is the peak brightness on the near side divided by that along the major axis. In addition, to examine the relative brightness of the two spiral arms, we define $R_{sp}$ as the ratio of the mean value in the radial range [0.18, 0.20]\arcsec of the near side profile, divided by the mean value in the radial range [0.22, 0.24]\arcsec on the far side profile. 
Each ratio is computed for both polarized intensity and the total intensity, and we use superscripts to indicate the radiation component, $R^{PI}$ for polarized intensity and $R^{I}$ for total intensity. 

These brightness ratios are sensitive to the dust scattering phase function and depend on the single scattering albedo and asymmetry parameters given the disk geometry \citep{Ma2022}. We find that $R^{I}$ is higher than $R^{PI}$ in all measured ratios, due to stronger forward scattering in total intensity. Both $R^{PI}_{N/F}$ and $R^{PI}_{N/M}$ decrease with wavelength, suggesting that the scattering becomes less anisotropic at longer wavelengths. This aligns with scattering grain sizes comparable to the observing wavelength. In contrast, the spiral brightness ratio $R^{PI}_{sp}$ remains approximately constant with wavelength, hinting at a different grain population or scattering regime in the spiral arms compared to the ring. 
These quantities and trends serve as diagnostic constraints in our later dust-scattering properties analysis. 

All $V$-band profiles show for the innermost part near $r\approx 100$~mas, inside the disk cavity, a substantial polarization signal. The contribution is much weaker for the $I'$-band, and hardly visible in near-IR. This indicates the presence of dust scattering inside the cavity, which is particularly strong at short wavelengths, and shows a bluer color than the dust in the rim. Notably, this polarized signal has not been detected before and is only revealed thanks to the high spatial resolution and short-wavelength coverage of the $V$-band observations.

\begin{table}[]
    \caption{Brightness ratios for all epochs. }
    \centering
    \begin{tabular}{c c c c c}
    \hline
    \hline
    Labels  & $R^{PI}_{N/F}$ & $R^{PI}_{N/M}$ & $R^{PI}_{\rm sp}$\\
    \hline
    V      & 1.33 $\pm$ 0.14 & 0.88 $\pm$ 0.14 & 1.02 $\pm$ 0.23 \\
    V-hop  & 1.49 $\pm$ 0.09 & 0.88 $\pm$ 0.04 & 1.32 $\pm$ 0.25\\
    I     & 1.30 $\pm$ 0.03 & 0.78 $\pm$ 0.03 & 1.05 $\pm$ 0.13\\
    J      & 1.08 $\pm$ 0.12 & 0.68 $\pm$ 0.05 & 0.97 $\pm$ 0.19 \\
    K      & 0.98 $\pm$ 0.03 & 0.65 $\pm$ 0.03 & 1.00 $\pm$ 0.09\\
    K-hop  & 0.95 $\pm$ 0.06 & 0.65 $\pm$ 0.06 & 0.95 $\pm$ 0.09\\
    \hline
             & $R^{I}_{N/F}$ & $R^{I}_{N/M}$ & $R^{I}_{\rm sp}$\\
    V-hop  & 1.84 $\pm$ 0.15 & 1.34 $\pm$ 0.34 &  2.68 $\pm$ 0.32\\
    K-hop  & 1.49 $\pm$ 0.27 & 1.54 $\pm$ 0.10 & 2.11 $\pm$ 0.20\\
    \hline
    \end{tabular}
    \label{tab:contrast}
\end{table}

\section{Model}
\label{sect: model}
Due to the complex substructure in HD~100453, various disk models have been developed, especially to study the disk-companion interaction \citep{Wagner2018, Gonzalez2020, Nealon2020}. While these hydrodynamic models reproduce ALMA dust emission and gas kinematics, they are not suited for a quantitative comparison with the observed polarized intensity and total intensity to characterize the dust scattering properties.  

To interpret the observed wavelength-dependent polarized intensity, total intensity, and degree of polarization, we adopt the parametric disk model from \cite{Benisty2017}, originally built with MCMax \citep{Min2009}. This model also successfully reproduces the system's SED, especially the outer disk’s infrared excess, providing a physically consistent dust density structure. We fully adopted the stellar, inner and outer disk parameters, and rebuilt the 3D dust structure with RADMC-3D \citep{Dullemond2012} without dust settling. This simplification is justified as the micron-sized and smaller grains responsible for scattered light are expected to be well coupled to the gas and remain suspended in the disk surface layers.
The NE and SW spiral arms parameterizations are slightly adjusted to better match the observed morphology (summarized in Table~\ref{tab: model-modified}). 
Radiative transfer simulations are then performed in full scattering mode, launching $10^6$ photon packets to produce polarized and total intensity images for the central wavelength of each filter. 
Although this model does not follow a self-consistent hydrodynamical evolution, and the spirals are implemented via artificial scale height enhancements, it offers the flexibility to match observed features in scattered light and allowed us to directly vary dust scattering properties.

\begin{figure*}
    \centering
    \includegraphics[width=0.95\linewidth]{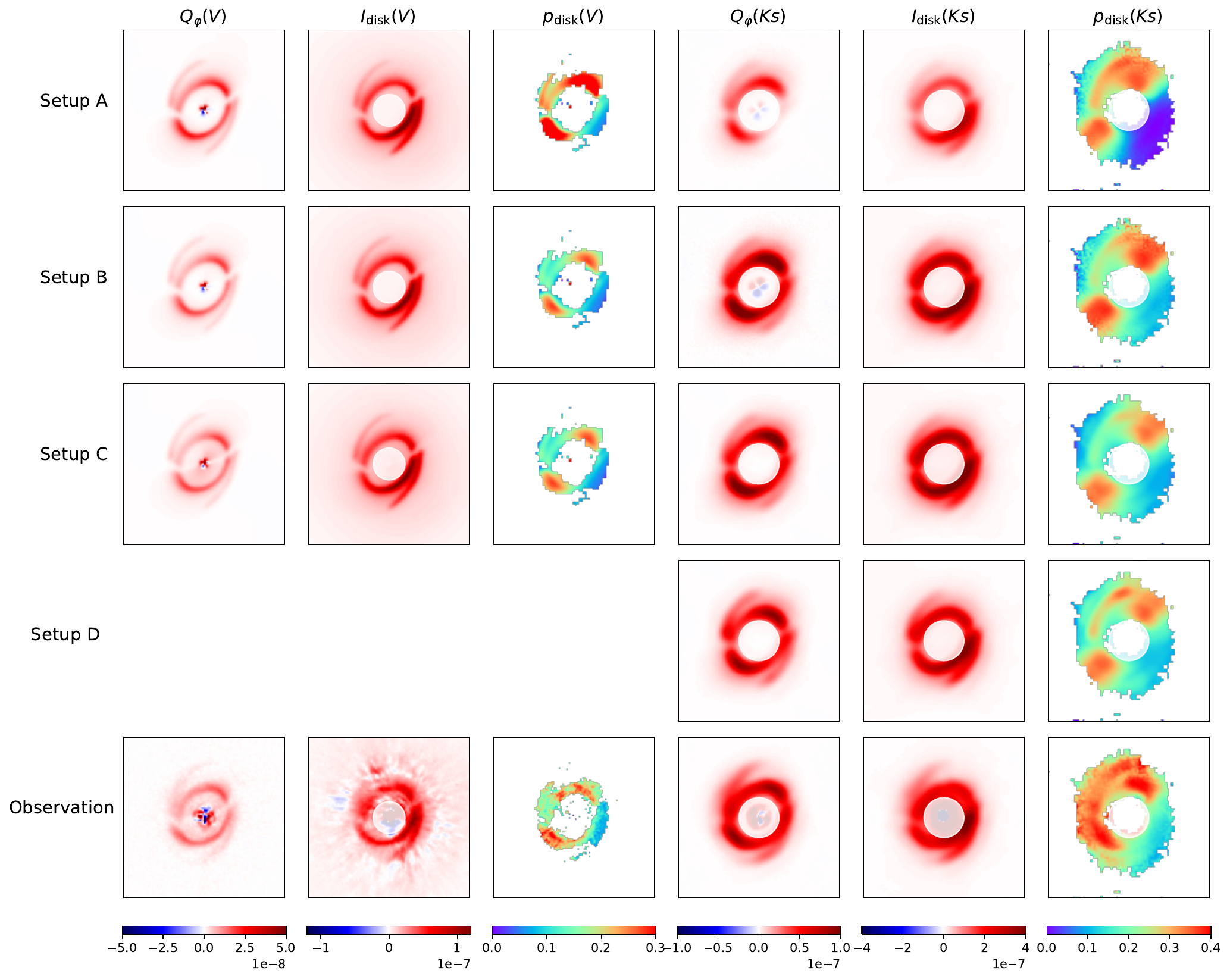}
    \caption{Simulated polarized disk intensity and total disk intensity in different setups using the best fitting parameters, convolved with corresponding PSFs. Setup A assumes single dust species DIANA 0.01-1$\mu m$ dust in both the inner and outer disk. Setup B has the same configuration as Setup A, but assumes single Henyey-Greenstein dust, with $(\omega, g, p_{\rm max}) = (0.32, 0.58, 0.37)$ at $V$-band and $(0.69, 0.53, 0.58)$ at $Ks$-band. Setup C introduces a new cavity zone between the inner and outer disk. The inner disk and the cavity contains DIANA 0.01-0.1$\mu m$ dust. The outer disk contains the same dust as in Setup B. Setup D has the same configuration as C but assumes different dust in the rim and the spiral in the outer disk for the $Ks$ band. ($\omega, g, p_{\rm max}) =(0.61, 0.47, 0.52)$ for the rim, and $(0.79, 0.49, 0.74)$ for the spirals. }
    \label{fig:model-v-k}
\end{figure*}

In this section, we start from the model as presented in \citet{Benisty2017} and explore modified setups informed by our observed constraints, aiming to characterize the dust properties in the disk. An overview of the simulated images for each setup is presented in Fig.~\ref{fig:model-v-k}, and we discuss them individually in the following subsections. 

\subsection{Setup A: DIANA standard dust} 
\label{sect: set-up-a}
This is a reproduction of \cite{Benisty2017} with minor adjustments to the geometric parameters of the spirals. The model assumes single dust species, the DIANA standard dust model, with the composition adopted from \citet{Woitke2016}, and a power-law size distribution $n(a)\propto a^{-3.5}$ between $a_{\rm min}=0.01\mu m$ and $a_{\rm max}=1 \mu m$ following \citet{Benisty2017}. The refractive index is calculated assuming the Bruggeman mixing rule. The opacity and the scattering matrix are calculated with a distribution of hollow spheres (DHS, \citet{Min2005}) approximation assuming $25\%$ porosity. 

This model was shown to reproduce the $Q_{\varphi}(x,y)$ observation at $I'$-band in \cite{Benisty2017}, and it roughly matches the $V$-band observation ( Figure~\ref{fig:model-v-k} second row), as expected for a not too different wavelength.
However, it fails to reproduce the $Ks$-band observations. Both the simulated polarized intensity $Q_{\varphi}$ and total intensity $I_{\rm disk}$ are significantly underestimated. Moreover, the simulated $Q_{\varphi}(x,y)$ lacks the near side, while the simulated $I_{\rm disk}(x,y)$ shows a too bright near side and too faint far side relative to the observation at $Ks$-band. 
The underpredicted brightness is due to the assumed dust's low albedo and strong forward scattering. The strong forward scattering also produces the too bright near side in simultated $I_{\rm disk}(x,y)$, while the deficit of polarization in low scattering angles causes the missing front side in the simulated $Q_{\varphi}(x,y)$. 

These discrepancies indicate that the DIANA standard dust is unlikely to dominate the surface layer of the disk. We tested variations in the maximum grain size, including both smaller and larger values up to several microns, but none produced satisfactory matches to the observations. To better constrain the dust properties without exhaustively exploring different compositions, size distributions, or porosities, we opt to characterize the dust via its scattering behavior. In the following setups, we define the scattering phase function using the Henyey-Greenstein formalism \citep{Henyey1941}, which allows for more direct control over the key scattering parameters.

\subsection{Setup B: Henyey-Greenstein dust approximation} 
\label{sect: set-up-b}
We introduce a user-defined dust species "HG dust", for which the scattering matrix is fully described by three parameters: the single scattering albedo $\omega$, the asymmetry parameter $g$, and the maximum degree of polarization $p_{\rm max}$. The Mueller matrix elements are defined as 
\begin{align}
    F_{11} &= \dfrac{\kappa_{scat}}{4\pi} \cdot \dfrac{1-g^2}{(1+g^2-2g\mu)^{3/2}} , \\
    F_{12} &= F_{11}\cdot p_{\rm max}\cdot \dfrac{\mu^2-1}{\mu^2+1} , \\
    F_{33} &= F_{11}\cdot \dfrac{2\mu}{\mu^2+1},  
\end{align}
where $\mu = \cos\theta_{scat}$. The $\kappa_{scat}$ is adopted from the DIANA standard model to preserve realistic scattering opacities. Here, $F_{11}$ represents the scattering phase function, which is the scattered intensity as a function of scattering angle. The element $-F_{12}$ is often referred to as the polarized scattering phase function. And $-F_{12}/F_{11}$ gives the degree of linear polarization, which is a symmetric bell-shaped profile centered at the scattering angle $90^\circ$, with its amplitude scaled by $p_{\rm max}$. We assume symmetric $F_{21}=F_{12}$, $F_{11}=F_{22}$ and $F_{33}=F_{44}$ which holds for spherical particles and for a collection of randomly oriented nonspherical ones. We neglect the circular polarization by setting all the other elements to 0. While $\omega$ does not appear explicitly in the scattering matrix element, it defines the absorption opacity $\kappa_{abs}$ via $\omega = \kappa_{sca}/(\kappa_{sca}+\kappa_{abs})$, required as an input to compute the radiative transfer.

We constructed a grid of 256 HG dust types by varying $\omega \in [0.10, 0.85]$ in steps of 0.05 and $g \in [0.10, 0.85]$ in steps of 0.05, fixing $p_{\rm max} = 1$ to reduce the grid size. For each HG dust, we computed polarized and total intensity images at the relevant wavelengths and convolved them with the corresponding PSF to compare with observations.

From the convolved images, we measured a set of key observables for each wavelength. 
For $V$ and $Ks$-bands where both $Q_{\varphi}(x,y)$ and $I_{\rm disk}(x,y)$ observations are available, we measure seven observables: integrated disk intensity $I_{\rm disk}/I_{\star}$, integrated polarized intensity $Q_{\varphi}/I_{\star}$, maximum polarization ${\rm max}(p_{\rm disk})$, and brightness ratios $R_{N/F}$ and $R_{N/M}$ for both polarized and total intensity. 

To facilitate fitting of scattering parameters, we interpolated the resulting observable grids to obtain continuous functions $I_{\rm disk}/I_{\star}(\omega, g)$, $Q_{\varphi}/I_{\star}(\omega, g)$
${\rm max}(p_{\rm disk})(\omega, g)$, and $R_{N/F}(\omega, g)$ and $R_{N/M}(\omega, g)$ for both polarized and total intensity. 
Particularly, since $Q_{\varphi}$ and ${\rm max}(p_{\rm disk})$ scale linearly with $p_{\rm max}$ to the 
first order approximation (when neglecting multiple scattering effects). 
We used the relation $Q_{\varphi}/I_{\star}(\omega, g, p_{\rm max}) = Q_{\varphi}/I_{\star}(\omega, g)*p_{\rm max}$ and similarly for ${\rm max}(p_{\rm disk})(\omega, g, p_{\rm max})$. The other observables do not depend on $p_{\rm max}$. 
We applied Markov Chain Monte Carlo (MCMC) fitting with flat priors in $\omega \in [0.10, 0.85], g\in [0.10, 0.85]$ and $p_{\rm max} \in (0,1)$ to maximize the log-likelihood for the seven observables in $V$ and $Ks$-bands using \texttt{emcee} \citep{Foreman-Mackey2013}. The best-fitting values are summarized in Table~\ref{tab: setup-b-param}, and the typical $1\sigma$-uncertainty is 0.03 for each parameter from MCMC fitting.  

We find that the brightness ratios are the determining factor for $g$. However, a single $g$ cannot simultaneously reproduce both $R^{PI}_{N/F}$ and $R^{PI}_{N/M}$ ratios measured from $V$-band observation. For example, using only five observables and considering only $R^{I}_{N/F}$ and $R^{PI}_{N/F}$ yields $(\omega, g, p_{\rm max}) = (0.27, 0.49, 0.37)$, and using only the $R_{N/M}$ gives $(\omega, g, p_{\rm max})=(0.49, 0.81, 0.42)$. This tension likely comes from the fact that the assumed HG phase function is a poor approximation at short wavelengths, where more complex scattering behaviors (e.g., skewed $-F_{12}/F_{11}$) are required. 
For the final $V$ and $K_s$-band results, we adopt the best-fit values from the full MCMC fitting and assign uncertainties based on the two individual fits to $R_{N/F}$ and $R_{N/M}$ (as summarized in Table.~\ref{tab: setup-b-param}). The smaller uncertainties at $K_s$-band indicate the HG approximation is more valid at longer wavelengths.

For $I'$ and $J$ bands in which only $Q_{\varphi}(x,y)$-observations are available, we measure three observables: the integrated polarized intensity $Q_{\varphi}/I_{\star}$, the brightness ratios for polarized intensity $R^{PI}_{N/F}$ and $R^{PI}_{N/M}$. We applied the same interpolation procedure, but only three observables are insufficient for a full parameter fit. 
We estimate $p_{\rm max}$ at $I'$ and $J$ by linear interpolation between $V$ and $Ks$, and fix $p_{\rm max}$ to fit $\omega$ and $g$ and the resulting values are given in Table~\ref{tab: setup-b-param}. The uncertainties are difficult to estimate for this procedure, but the goal is to have rough parameters for the simulations of $Q_{\varphi}(x,y)$ maps for the later analysis on the cavity flux.

The best-fit convolved simulations are shown in the second row of Fig.~\ref{fig:model-v-k}, matching the observed images well in both $V$ and $Ks$ bands as we fit them separately. 
This setup demonstrates that the multiwavelength observations can be reproduced using Henyey-Greenstein scattering approximation. The fitted wavelength-dependent scattering trend provides key constraints on dust properties and hints at sub-micron-sized compact aggregates as the dominant scatterers on the disk surface, which we explore in more detail in Section~\ref{sect: dust-properties}.
However, Set-up B cannot account for two features in the observations: the presence of polarized flux within the cavity at the $V$-band, which cannot be explained by PSF smearing alone, and the insufficient brightness of the spirals in the $K_s$-band. These discrepancies motivate the targeted modifications introduced in Set-up C and D.

\begin{table}[]
    \centering
    \caption{Fitting results for the dust scattering parameters for each wavelength.}
    \label{tab: setup-b-param}
    \begin{tabular}{l c c c }
    \hline
    \hline
     Filters   &  $\omega$ & $g$ & $p_{\rm max}$ \\
    \hline
    $V$ &  $0.32^{+0.04}_{-0.05}$ & $0.58^{+0.23}_{-0.09}$ & $0.37^{+0.05}_{-0.02}$ \\
    $Ks$ & $0.69^{+0.03}_{-0.03}$ & $0.53^{+0.04}_{-0.07}$ & $0.58^{+0.03}_{-0.03}$ \\
    \hline
    $I'$ & 0.42 & 0.55 & 0.40\tablefootmark{a} \\
    $J$ & 0.61 & 0.51 & 0.45\tablefootmark{a} \\
    \hline
    \end{tabular}
    \tablefoot{
    \tablefoottext{a}{Estimated by linear interpolating $p_{\rm max}(\lambda)$ between $V$ and $Ks$ and using a fixed $p_{\rm max}$} in the fitting. 
    }
\end{table}

\subsection{Setup C: New zone for cavity}
\label{sect: set-up-c}
To explain the $V$-band flux detected inside the cavity, which is not due to the convolution effect alone, we introduce a cavity zone to the model. The cavity is assumed to be geometrically aligned with the outer disk, supported by the inner-disk shadow matching the expected orientation in the outer ring; and optically thin with $M_{\rm dust} = 1.5 \times 10^{-10} M_{\odot}$, derived by adjusting the dust mass in RT model to match the observed $V$-band polarized flux. 

The dust surface density in the cavity is described by
$\Sigma(r) \propto r^{-\epsilon}\exp[(r/R_{\rm tap})^{\epsilon-2}]$ with $\epsilon = -1$
for a nearly flat surface density profile that gently decreases inward (see Fig.~\ref{fig:surface-density}).
This choice is motivated by hydrodynamic simulations of the planet–disk interactions, where material inflow crosses a planet-carved gap and feeds the star, leading to a relatively uniform surface density inside the planet's orbit \citep[see Fig. 5 of][]{Tatulli2011}. Our adopted $\epsilon = -1$ captures this behavior, and we find that small variations in $\epsilon$ have little impact on our modeling results. 

Adding the optically thin dust in the cavity does not affect the outer disk appearance in the scattered light, so we fixed the outer disk parameters to those derived in Setup B (Table.~\ref{tab: setup-b-param} in this analysis. Since the inner disk is optically thick and unresolved, the dust species in the inner disk cannot be constrained so we assumed the same dust species as in the cavity. 

In the optically thin cavity, the scattered light intensity is directly proportional to the scattering opacity $\kappa_{\rm sca}$ of the dust, which is primarily sensitive to the grain size distribution \citep{Bohren1983}. Therefore, the wavelength dependence of the cavity flux encodes information about dust size. 
However, we cannot reliably extract the total intensity in the cavity or apply PSF correction on $Q^{\rm cav}_{\varphi}/I_{\star}$ due to the faintness of the cavity signal and its proximity to the star. 
Instead, we adopt the forward-modeling approach: using the DIANA standard dust as a reference, we fix all parameters except for $a_{\rm max}$, and compare the convolved simulated polarized intensity $Q^{\rm cav}_{\varphi}/I_{\star}$ to the observed value. This allowed us to simultaneously constrain the cavity dust properties and account for PSF smearing effects. 

\begin{figure}
    \centering
    \includegraphics[width=0.95\linewidth]
        {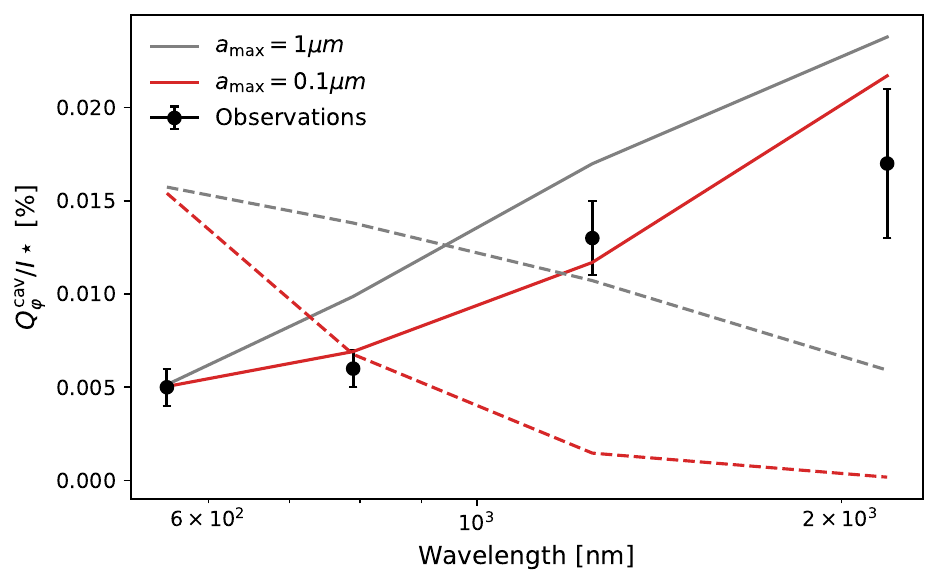}
    \caption{Polarized flux integrated in the cavity region $Q^{\rm cav}_{\varphi}/I_{\star}$. The black scatter points are measurements from observations (Table.~\ref{tab: polarised-flux}). The solid gray and red lines are the contrasts measured from convolved simulations assuming DIANA $a_{\rm max}=1$ and $0.1\mu m$ in the cavity respectively. The dashed gray and red lines are measured from simulations before convolution.}
    \label{fig: cavity-flux}
\end{figure}

\begin{figure}
    \centering
    \includegraphics[width=0.98\linewidth]{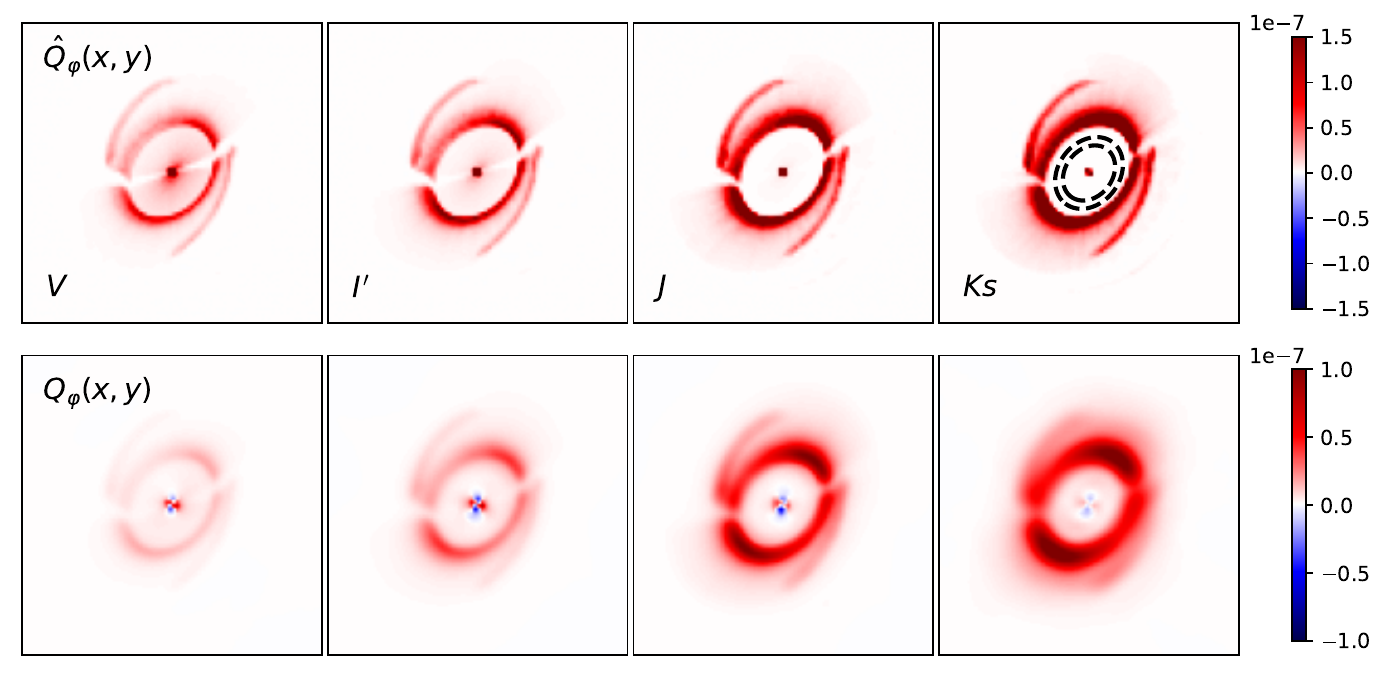}
    \caption{Simulated polarized intensity maps in Setup C. Top row: Intrinsic $\hat{Q}_{\varphi}(x,y)$ without convolution. Bottom row: $Q_{\varphi}(x,y)$ after convolution with the corresponding PSF at $V$, $I'$, $J$, and $Ks$. The dust parameters in the outer disk are taken from Table.~\ref{tab: setup-b-param}. The inner disk and cavity assume DIANA dust with $a_{\rm max}=0.1\mu m$. The dashed lines in $\hat{Q}_{\varphi}(x,y)$-image at $Ks$ band encloses the cavity region. }
    \label{fig: setup-c-qphi}
\end{figure}

As shown in Fig.~\ref{fig: cavity-flux}, the intrinsic $\hat{Q}^{\rm cav}_{\varphi}/I_{\star}$ (gray dashed line) decreases with wavelength assuming DIANA standard dust. 
After convolution, the $Q^{\rm cav}_{\varphi}/I_{\star}$ increases with wavelength (solid gray lines) due to the PSF smearing effect, however with a steeper gradient than the measurements. 
This mismatch indicates an even stronger wavelength dependence, i.e., a faster drop in the $\kappa_{\rm sca}(\lambda)$, is required to explain the observed cavity intensity. Such a trend can be achieved by reducing the maximum size, $a_{\rm max}$, in the size distribution, shifting the dust population closer to the Rayleigh regime $\kappa_{\rm sca}(\lambda) \propto \lambda^{-4}$ if the wavelength dependence of the refractive index is negligible. A larger $a_{\rm max}$ ($>1\mu m$) results in an even steeper gradient in $Q_{\varphi}^{\rm cav}/I_{\star}(\lambda)$ and can therefore be excluded. We found that reducing $a_{\rm max}$ to $0.1\mu m$ (red lines in Fig.~\ref{fig: cavity-flux}) is enough to produce the observed trend. The simulated $Q_{\varphi}(x,y)$ maps for each wavelength are shown in Fig.~\ref{fig: setup-c-qphi}. We see that in the intrinsic polarization, the cavity becomes fainter with wavelength. And the $Q^{\rm cav}_{\varphi}/I_{\star}$ at $Ks$-band is mostly attributed to convolution effect. 

The simulations with $a_{\rm max}=0.1\mu m$ dust in the cavity are shown in Fig.~\ref{fig:model-v-k} in the third row, reproducing the cavity signal.  
Even steeper spectral slopes can be realized by lowering $a_{\rm max}$ further or by steepening the size distribution slope ($p>3.5$), which makes the effective dust size even smaller. In any case, our results indicate that the scattering grains in the cavity cannot be larger than 0.1$\mu m$. This is consistent with theoretical predictions of dust filtration, where large grains are trapped at the outer edge of a planet-induced gap and only small grains can flow inward \citep{Zhu2012}. 
 
\subsection{Setup D: Different dust species in spirals}
In the previous setups, we assumed a uniform dust population across the outer disk. This is certainly an oversimplification, especially for a structured system such as HD~100453, where features such as the bright ring and trailing spirals may host different dust populations. 
Specifically, in the convolved $Q_{\varphi}$-image, the spiral flux is underestimated by $\sim$20\%. In total intensity, the signal in spiral is reproduced by over-predicting the rim flux by $\sim$10\%. These discrepancies suggest a potential variation in dust properties.
Therefore in Setup D, we explore this possibility by separating the dust population in the outer disk.
We fix the inner disk and cavity components as in Setup C. We fit the HG dust parameters independently for the rim and the spirals, following a similar process as in Setup B. 

For the rim, we chose seven observables: the integrated flux $Q^{\rm rim}_{\varphi}/I_{\star}$, $I^{\rm rim}_{\rm disk}/I_{\star}$, the maximum polarization ${\rm max}(p_{\rm disk})$ and the brightness ratios $R_{N/F}$ and $R_{N/M}$ for both polarized and total intensity. The ${\rm max}(p_{\rm disk})$, $R_{N/F}$, and $R_{N/M}$ are the same as measured for the whole disk. 
For the spiral, we chose five observables: $Q^{\rm sp}_{\varphi}/I_{\star}$, $I^{\rm sp}_{\rm disk}/I_{\star}$, ${\rm max}(p^{\rm sp}_{\rm disk})$ and the brightness ratio $R^{I}_{sp}$ and $R^{PI}_{sp}$. The $R_{sp}$ acts as a near-far brightness ratio for the spiral, while the near side to major axis ratio is not available. The independent fitting applies only for the $Ks$-band.
This is not feasible for the $V$-band data because the spirals conincide with the strong speckle ring of the PSF and this noise introduces large uncertainties for the measured polarization of the spirals. 

At $Ks$-band, the best-fitting parameters are $(\omega, g, p_{\rm max}) = (0.61\pm0.03, 0.47\pm 0.02, 0.52\pm0.04)$ for the rim, and $(0.79\pm0.02, 0.49\pm0.04, 0.74\pm0.06)$ for the spiral. The uncertainty represents $1\sigma$ deviation for the confidence interval of MCMC fitting. The convolved best-fitting model images are shown in Fig.~\ref{fig:model-v-k} row Setup D.
These results confirm that distinct dust properties in the spirals can better reproduce their brightness at $Ks$ band. Compared to the global-fit parameters from Setup C, the rim values change only slightly, since the rim dominates the disk’s total brightness. The spiral region, however, favors higher $\omega$ and $p_{\rm max}$ values, suggesting more reflective grains with higher intrinsic polarization. This finding is discussed further in Section~\ref{sect: dust-properties}

\section{Discussion}
\label{sect: discussion}
\subsection{Intrinsic disk polarization parameters}
While the intrinsic disk-integrated azimuthal polarization $\hat{Q}_\phi/I_\star$ can be obtained directly from the observed data using a PSF-based correction map (Sect.~3.2), other parameters cannot be corrected in the same way because the convolution effects depend in a more complex way on the detailed structure of the disk.
The RT modeling provides a high-quality nonconvolved simulation of both the disk total and polarized scattered light images, enabling accurate measurements of intrinsic values. In the following analysis, we adopt the best-fit models from Setup C for $V$-band and Setup D for $Ks$-band. 

The full-disk intensity at $Ks$ is not available from the observation due to the coronagraph. From model Setup D, we integrated the disk intensity and obtained $I_{\rm disk}/I_{\star} = 4.2 \pm 0.6\%$ when integrating the convolved disk intensity in the elliptical region, and $4.8 \pm 0.7\%$ when integrating the intrinsic model. 
This difference arises because the PSF convolution spreads part of the disk signal into the central 0.125\arcsec region, which lies inside the inner working angle and is excluded from the integration.
Similarly for $V$-band, we integrated the unconvolved model and obtained $I_{\rm disk}/I_{\star} = 1.5\%$. 
The uncertainty is assumed the same as for the coronagraphic image. 

Previous measurements of the total intensity for HD~100453 have employed advanced imaging techniques and provided valuable measurements for comparison. For example, \citet{Long2017} reported disk-integrated fluxes of $I_{\rm disk}/I_{\star}=1.8\pm 0.1\%$ at $J$ and $2.5\pm 0.2\%$ at $H$ band using angular differential imaging (ADI) with KLIP \citep{Soummer2012}, which was among the first efforts to quantify the integrated scattered light in this system. While ADI techniques are known to introduce some level of self-subtraction for extended disk structures, these values fall neatly between our $V$- and $Ks$-band measurements, supporting a broadly consistent trend in the scattered light intensity across wavelengths.
The K-hop data were presented by \citet{Ren2023} and high-quality extracted disk intensity image using NMF data imputation \citep{Ren2020} was made publically available. Although the integrated disk flux was not reported in their work, we were able to re-analyze the image, correct for the coronagraphic transmission, and estimate $I_{\rm disk}/I_{\star} = 3.4\%$, which only differs from our measurement by 10~\%. This is likely due to the inclusion of companion during PSF subtraction. 
Our improved flux estimates allow for more accurate and consistent measurements across filters, supporting further reliable color analysis. 

\begin{table}[]
    \centering
    \caption{Measurements for observed and intrinsic disk radiation parameters for the $V$ and $Ks$ bands.}
    \label{tab:revised_measurements}
    \resizebox{\linewidth}{!}{%
    \begin{tabular}{c c c c c c c}
    \hline
    \hline
    Filters & $I_{\rm disk}/I_{\star}$ & $I_{\rm disk}^{\rm rim}/I_{\star}$ & $I_{\rm disk}^{\rm sp}/I_{\star}$ & ${\rm max}(p_{\rm disk})$ & $\langle p_{\rm disk} \rangle$ \\
     & \% & \% & \% & \% \\
    \hline
    \noalign{Observed disk parameters}
    $V$ & $1.2\pm0.3$ & - & - & $28\pm2$ & $16\pm5$ \\
    $Ks$ & $3.8\pm0.5$ & - & - & $39\pm4$ & $33\pm6$ \\
    \noalign{Intrinsic disk parameters}
    $V$  &  $1.5\pm 0.5$  & $1.2\pm0.4$ & $0.3\pm 0.1$ & $40\pm 3$ & $20\pm 6$\\
    $Ks$ &  $4.8\pm0.7$   & $3.5\pm 0.5$ & $1.3\pm0.2$ & $55\pm 6$ & $25\pm 5$\\
    \hline
    \end{tabular}
    }
\end{table}

Using the best-fit models at $V$ (Setup C) and $Ks$ (Setup D), we derive convolution-corrected fluxes for the rim and spirals separately in the concentric elliptical aperture, and the results are given in Table~\ref{tab:revised_measurements}. The uncertainties are adopted from observed measurements. 
We also derived the corresponding degree of polarization maps and determined the intrinsic maximum polarization ${\rm max}(\hat{p}_{\rm disk})$. We find $40\%$ at $V$ and $55\%$ at $Ks$-bands, higher than $28\%$ and $39\%$ earlier derived from the observed $Q_{\varphi}$ and $I_{\rm disk}$ maps directly. Since our convolved model reproduces the observed values well, this degradation of maximum degree of polarization must be attributed to the PSF convolution. The convolution smears the intensity and polarization, and further reduces the polarization signal by the mutual cancellation of the positive and negative components of the Stokes $Q$ and $U$ maps \citep{Schmid2025}. While the disk-averaged degree of polarization $\langle p_{\rm disk}\rangle$ at $Ks$ is higher before correction, because the coronagraph partly hides the front side where the degree of polarization is low. 

\paragraph{Color measurements}
With these improved intrinsic flux measurements, we characterize the spectral dependence of disk reflectivity using the logarithmic color gradient $\eta_{\lambda_1/\lambda_2} = (\log L_{\rm disk, \lambda_2} - \log L_{\rm disk, \lambda_1})/{\log(\lambda_2/\lambda_1)}$, 
where $L_{\rm disk}$ corresponds to $\hat{Q}_{\varphi}/I_{\star}$ for $\eta^{PI}$ and $I_{\rm disk}/I_{\star}$ for $\eta^{I}$ \citep{Tazaki2019}.
The colors are classified into blue ($\eta<-0.5$), gray ($-0.5<\eta<0.5$), and red ($\eta>0.5$). For the HD~100453, we measure red color across $V$ to $Ks$ band, with a disk-integrated $\eta^{PI}_{V/K}=1.04 \pm 0.04$. More specifically, we find red colors both within the visible range $\eta^{PI}_{V/I'} = 1.19 \pm 0.07$ and from visible to near-infrared $\eta^{PI}_{I'/J} = 1.53 \pm 0.21$, consistent with the earlier estimation from \citet{Ma2024}, with significantly improved precision. 
Additionally in this work, we measure reddish-gray color in the infrared $\eta^{PI}_{J/K} = 0.55 \pm 0.23$. 
When analyzing disk substructures, the rim shows a similar color to the full disk $\eta^{\rm PI, rim}_{V/K} = 1.00\pm 0.01$, while the spirals appear even redder, with $\eta^{\rm PI, sp}_{V/K} = 1.41 \pm 0.04$. This redder trend in the spirals persists across other wavelength pairs as well. 

Also for the total intensity, we estimate a red color $\eta^{I}_{V/K} = 0.85 \pm 0.28$ for the full disk. 
For substructures, we estimate $\eta^{\rm I,rim}_{V/K} = 0.8 \pm 0.3$ and $\eta^{\rm I,sp}_{V/K} = 1.1 \pm 0.3$, consistent with the full disk. \citet{Long2017} previously reported a bluer color in the spirals and interpreted this as evidence for smaller grains. While their result was based on ADI with KLIP, which is known to introduce self-subtraction biases in extended structures, our analysis yields a consistent red gradient across the disk. In particular, we find the spirals to be at least as red as the rim, if not redder. This suggests that the dust in the spirals is not necessarily smaller and could instead be shaped by other factors such as porosity or viewing geometry, which we'll discuss in the following. 

\subsection{Dust species}
\label{sect: dust-properties}
\begin{figure}
    \centering
    \includegraphics[width=\linewidth]{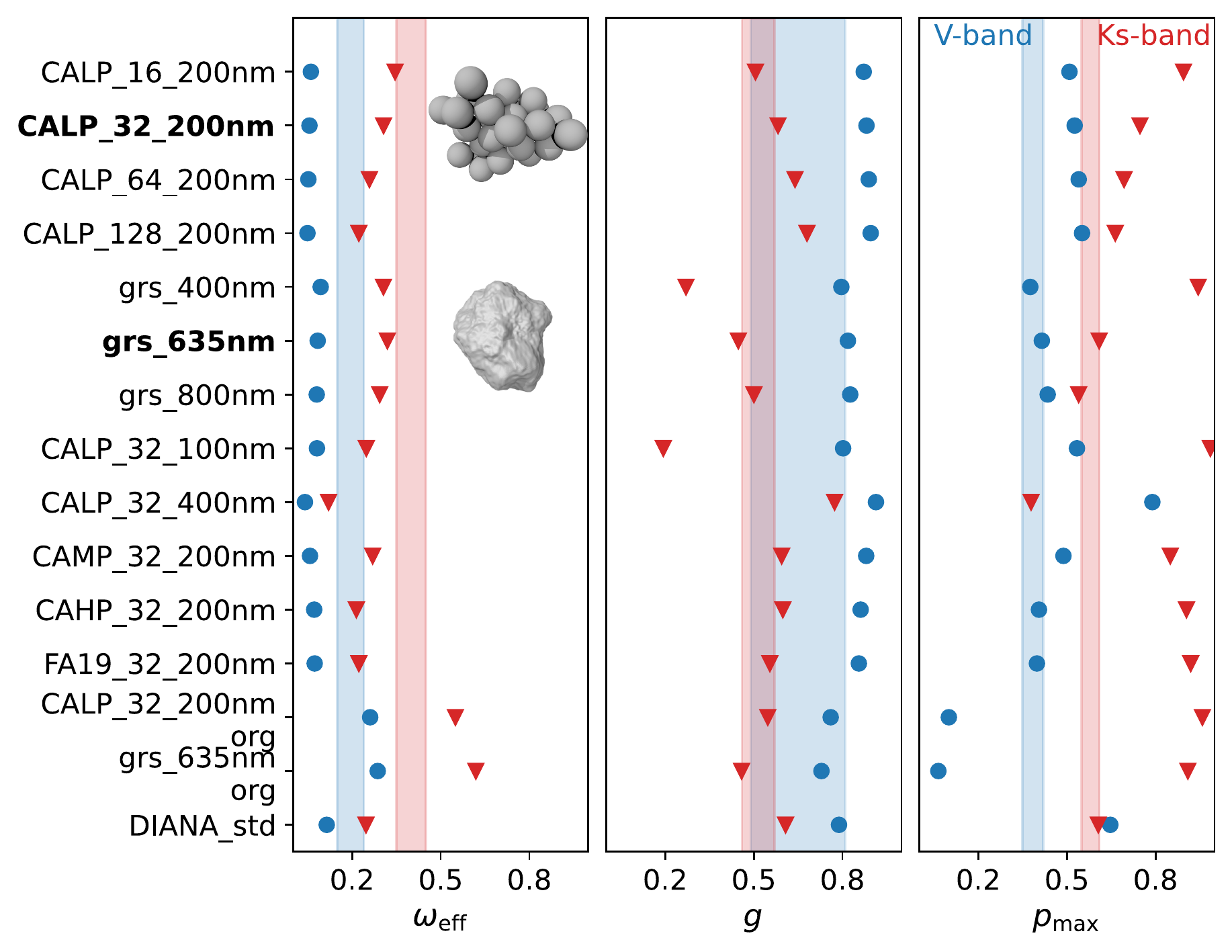}
    \caption{Comparison of the scattering properties $\omega_{\rm eff}$, $g$, and $p_{\rm max}$ of selected dusts species from \texttt{AggScatVIR} database. The dusts are labeled by morphology\_number of monomers\_monomer size. The morphology includes compact aggregates CALP (low porosity), CAMP(medium porosity), and CAHP (high porosity) and fractal aggregates FA19 ($d_f$=1.9), further characterized by the number of monomers and monomer sizes in nanometers. Irregular grains (GRS) are characterized by volume equivalent radii. If not specified, the composition \texttt{amc} is used. CALP\_32\_200nm and grs\_635nm examples are placed to show their morphology. }
    \label{fig: compare-dust}
\end{figure}

\begin{figure}
    \centering
    \includegraphics[width=0.95\linewidth]{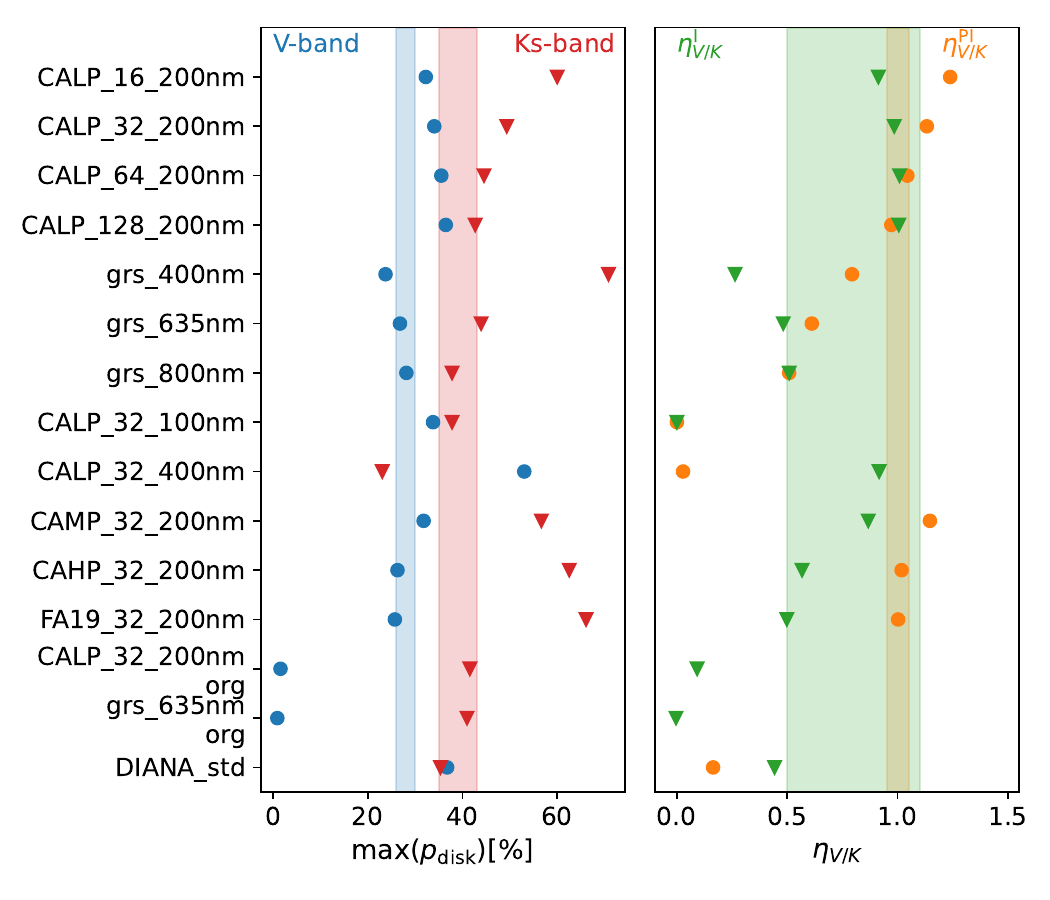}
    \caption{Comparison of $\max(p_{\rm disk})$(left column) and disk-integrated colors $\eta_{V/K}$(right column) when assuming different dust species from \texttt{AggScatVIR} database. The label convention is the same as in Fig.~\ref{fig: compare-dust}. }
    \label{fig: compare-pdisk}
\end{figure}

The scattering parameters derived from forward model fitting assuming Henyey-Greenstein scattering show strong wavelength dependence from $V$ to $Ks$ bands. 
Specifically, the single scattering albedo $\omega$ increases from about 0.3 to 0.7, the asymmetry parameter remains around 0.5, and the maximum polarization $p_{\rm max}$ increases from about 0.4 to 0.6. 
These trends enable us to constrain the grain structure and composition on the disk surface. 

\subsubsection{Constraints with aggregate models}
To interpret the derived scattering parameters in terms of dust structure, we turn to physically motivated aggregate dust models. We use the \texttt{AggScatVIR} (Aggregate Scattering for Visible and Infrared wavelengths) database \citep{Tazaki2023}, which provides a comprehensive library of light-scattering properties for aggregates with varying porosity, monomer size, and number of monomers, based on realistic collisional growth scenarios. 
Two compositions are considered: \texttt{amc}, a mixture of pyroxene silicate, water ice, troilite, and absorbent carbon; and \texttt{org}, which replaces the absorbent carbon with reflective carbon.

For all 360 models with power-law aggregate size distribution in \texttt{AggScatVIR} database, we calculated their scattering parameters $(\omega, g, p_{\rm max})$.
We begin by filtering and identifying dust models that show a moderate increase in $p_{\rm max}$ from $V$ to $Ks$ bands, as seen in our data. We focus on $p_{\rm max}$ because it is particularly sensitive to the dust structure and has been shown to strongly constain grain porosity and aggregate types \citep{Tazaki2022, Tazaki2023}. The filtering points to two promising \texttt{amc} families: CALP (compact aggregate with low porosity) models with 200 nm monomers and GRS (irregular grain).

To further investigate how individual dust properties influence the wavelength-dependent scattering behavior, we selected 14 representative models centered on these two families. In addition to the best-matching CALP and GRS models, we include variants where a single parameter—monomer size, porosity, or number of monomers—is altered. DIANA standard dust model used in Setup B is also included for comparison. Fig.~\ref{fig: compare-dust} presents their scattering properties, where models are labeled as aggregate type\_number of monomers\_monomer size. The shaded region indicates the HG parameter ranges derived from our modeling (Table 5).

As shown in the third column of Fig.~\ref{fig: compare-dust}, irregular grains of volume equivalent radii $a_V \approx 0.6-0.8~\mu m$ reproduce the observed increase in $p_{\rm max}$ the best. Compact aggregates composed of 200~nm monomers with similar volume equivalent radii $a_V$ also reproduce the trend. In contrast, more porous aggregates (e.g., CAMP, CAHP, and FA19, with porosity $\mathcal{P} > 70\%$) or smaller monomer size (e.g., CALP\_32\_100nm) tend to overpredict the $p_{\rm max}(\lambda)$ gradient, and $p_{\rm max}$ value in the near-infrared, which is inconsistent with our observations. Larger dust $a_V>1~\mu m$ does not show $p_{\rm max}$ increase. 

Secondly, to evaluate if the candidates can also reproduce the observed
red color, we adopted the effective albedo as defined in \citet{Mulders2013} to account for the inclined disk geometry: 
\begin{equation}
    \omega_{\rm eff} = \dfrac{2\omega\int_{\theta_1}^{\theta_2}F_{11}(\theta)\sin\theta d\theta}{(\cos\theta_1 -\cos\theta_2)\int_0^{\pi}F_{11}(\theta)\sin\theta d\theta},
\end{equation}
where $\theta$ represents the scattering angle, and the integration limits are set to $[\theta_1, \theta_2] = [50^\circ, 130^\circ]$ in this study. The results are shown in the first column of Fig.~\ref{fig: compare-dust}. 

While we did not search to fit the absolute $\omega_{\rm eff}$ values because the database has only two possible compositions, 
the increasing trend of $\omega_{\rm eff}(\lambda)$ is encouraging, as it suggests these aggregates can produce the observed red color in both the polarized and total disk reflectivity. 
The absolute value of $\omega$ mainly depends on the dust composition. In our models, the absorptive carbon species (\texttt{amc}) gives a reasonable match, though slightly increasing the fraction of more reflective components such as silicates and ices could improve the fit to the measurements. Still, what we really constrain is the refractive index, not the exact composition. Therefore, combinations with similar refractive indices as \texttt{amc} could produce the same scattering behavior.

Regarding the asymmetry parameter $g$, all models show a decrease with wavelength, consistent with our observations, though $g(V)$ is generally higher than the HG fit. The volume-equivalent grain size appears to be the most critical factor, as models such as CALP\_32\_200nm and grs\_635nm reproduce $g$ values at both bands effectively. Increasing the grain size slightly (e.g., from 32\_200nm to 128\_200nm or from grs\_400nm to grs\_800nm) results in higher $g(Ks)$ values, leading to overly forward-scattering behavior in the $Ks$ band. Sub-micron-sized dust give a marginally good match, while porosity and composition appear to have only secondary influence. 

We also note that our observational constraints on $g$ are limited by the modest inclination of the system, which restricts us to a narrow range of scattering angles near $90^\circ$. This can bias estimates of $g$ derived from the observed brightness contrast (see discussion in \citet{Hughes2018} for a similar issue in debris disks). While our modeling captures general trends, we caution that redder colors could also arise from more isotropic scattering behavior or other geometric effects.

The \texttt{AggScatVIR} scattering properties serve as a first diagnostic tool, giving us insight into the types of aggregate structures that could explain our measured scattering parameters. As a final test, we validate these candidate aggregates by directly inserting them into RT simulations, comparing the predicted total disk color and degree of polarization against observations (Fig.~\ref{fig: compare-pdisk}). We find that CALP and GRS reproduce both red color and degree of polarization. By contrast, aggregates with higher porosity ($\mathcal{P}$>70\%) can reproduce the red color, but overestimate the degree of polarization. 

In summary, combining the initial filtering from the \texttt{AggScatVIR} database with full RT modeling, we conclude that the scattering properties in HD100453’s surface layers are best explained by compact, low-porosity aggregates or irregular grains with volume-equivalent sizes around $0.8\mu m$.

Importantly, HD 100453 is a gas-depleted disk, with gas-to-dust ratios of 15–45 \citep{vanderplas2019, Collins2009}, which favors the existence of compact aggregates. Our results suggesting sub-micron, compact aggregates in the surface layer, could be interpreted as the evidence that the grains have undergone some dust processing. In contrast to gas-rich disks such as IM Lup, where fluffy, primordial aggregates are inferred \citep{Tazaki2023}, dust in HD 100453 appears more evolved. The low porosity might arise from collisional compaction or fragmentation in the low-gas environment \citep{Tanaka2023, Michoulier2024}. Additionally, the absence of large grains in the surface layer could reflect vertical settling in a disk with moderate turbulence, consistent with expectations for gas-depleted systems. 

\subsubsection{Substructure-dependent dust scattering behavior}
The origin of the spirals in HD~100453 has long been debated, with popular explanations including shadows cast by a misaligned inner disk \citep{Benisty2017} and tidal perturbations from the stellar companion\citep{Wagner2018, vanderplas2019}. While both scenarios are plausible, recent kinematic measurements using high-resolution ALMA and SPHERE imaging favor the companion-driven origin \citep{Rosotti2020, Xie2023}. However, how these spirals influence dust evolution remains unclear. Existing simulations explored dust dynamics in both scenarios \citep{Cuello2019, Gonzalez2020, Nealon2020}, but they were not designed to interpret the polarimetric scattered light images. The simulations followed the gas and dust evolution and predicted grain size distributions, but the dust is typically treated with Mie theory assuming astrosilicates. 
Introducing more realistic grains, especially nonspherical grains, would improve future comparisons with observations to study how dust evolution varies across substructures. 

Our analysis reveals a higher maximum degree of polarization $p_{\rm max}$ in spirals than the main ring at $Ks$ band, suggesting dust grains in the spirals are more porous or smaller. This trend is robust against geometric uncertainties, as $p_{\rm max}$ is insensitive to moderate changes in scattering surface height. To test the effect of geometry, we varied the spiral scale height in the RT model. Increasing the height modestly shifts the best-fit scattering parameters (e.g., increasing both $\omega$ and $g$), but does not affect $p_{\rm max}$ significantly. Therefore, while the absolute values may depend on geometry, the relative enhancement of $p_{\rm max}$ in the spirals is likely intrinsic to the dust. 
The elevated $p_{\rm max}$ in the spirals is consistent with compact aggregates of lower porosity in the ring and slightly more porous grains in the spirals. These findings highlight the potential of spiral arms to show differential grain growth, waiting for further investigation with physically consistent grain evolution models.

The cavity region also shows a distinctly different dust population. Our forward modeling indicates that only small grains ($a_{\rm max} \lesssim 0.1\mu$m) can reproduce the observed steep drop in polarized flux with wavelength. This supports a scenario where large grains are filtered out at the cavity edge, consistent with theoretical expectations of planet-induced dust filtration \citep{Zhu2012, Pinilla2016}, where the gas is still present \citep{vanderplas2019}. These results highlight the potential to reveal a small grain population and provide new evidence of dust filtration in other transition disks with gas-filled cavities (e.g., HD~135344B, \citet{Carmona2014}) by including short-wavelength PDI observations in future studies. 

\section{Conclusions}
\label{sect: conclusion}
We have presented new polarimetric observations of HD~100453 at the $V$, $I'$, $J$, and $Ks$ bands. We considered the PSF convolution effects and corrected for the cancellation in $Q_{\varphi}$-images to measure the intrinsic, integrated polarized flux. We performed reference differential imaging in star-hopping observing sequences at the $V$ and $Ks$ bands to extract total disk intensity images. We demonstrated that reference differential imaging is also applicable to ZIMPOL observations in the $V$-band, and we improved the disk intensity extraction at the $Ks$-band. We measured the integrated disk intensity, $I_{\rm disk}$, and derived a red color. We recognized an intrinsic scattering signal in the cavity in $V$ and $I'$ images. 
We also measured a brightness ratio between the near and far side, which decreases with wavelength.

We then characterized the dust properties using the radiative transfer RADMC-3D disk model with four different setups. Reproducing the simulation of \citet{Benisty2017} using the DIANA standard dust model revealed a deficit on the total and polarized disk flux in the $Ks$-band observations.
Adopting Henyey-Greenstein scattering phase functions for the dust reproduced the observation at both wavelengths, with best-fitting scattering parameters $(\omega, g, p_{\rm max})$ varying from (0.32, 0.58, 0.37) at $V$ to (0.69, 0.53, 0.58) at $Ks$. The dust dominating the upper layer of the disk is most likely composed of sub-micron-sized low-porosity aggregates and/or irregular grains. Introducing dust in the cavity zone between the inner and outer disks, we characterized the maximum dust size, $a_{\rm max}<0.1\mu m$, by fitting the cavity flux as a function of wavelength. Finally, we found that the dust in the spirals produces a higher scattering polarization, $p_{\rm max}$, when compared to regions further in, hinting at more porous or smaller dust.  

The results pointing to different dust species across the disk reflect distinct dust formation and evolution processes, such as grain growth in spirals versus the main ring and dust filtration at cavity edges. Linking these properties to specific mechanisms offers valuable insight into planet formation. Future studies that will expand the sample to a larger number of disks will be essential for identifying common patterns and constraining the physical processes shaping protoplanetary environments.

\section*{Data availability}
The reduced $Q_{\varphi}$ images in Fig.~\ref{fig: observations} and extracted disk intensity $I_{\rm disk}$ maps in Fig.~\ref{fig:disk-intensity-vk} are are available at the CDS via \url{https://cdsarc.cds.unistra.fr/viz-bin/cat/J/A+A/702/A78}.

\begin{acknowledgement}
We thank the anonymous referee for the valuable comments
on the manuscript. We thank Andrés Carmona González for helpful discussions on the surface density setups. We are also grateful to Bin Ren for his input on disk intensity extraction, and to Anthony Boccaletti for providing the coronagraphic transmission profiles.
J.M. thanks the Swiss National Science Foundation for financial support under grant number P500PT\_222298. R.T. is supported by JSPS KAKENHI Grant Numbers JP25K07351. G.D. and F.M. acknowledge funding from the European Research Council (ERC) under the European Union's Horizon Europe research and innovation program (grant agreement No. 101053020, project Dust2Planets.)
\end{acknowledgement}

\bibliographystyle{aa} 
\bibliography{biblio}      

\begin{thebibliography}{52}
\expandafter\ifx\csname natexlab\endcsname\relax\def\natexlab#1{#1}\fi

\bibitem[{{Benisty} {et~al.}(2023){Benisty}, {Dominik}, {Follette}, {Garufi}, {Ginski}, {Hashimoto}, {Keppler}, {Kley}, \& {Monnier}}]{Benisty2023}
{Benisty}, M., {Dominik}, C., {Follette}, K., {et~al.} 2023, in Astronomical Society of the Pacific Conference Series, Vol. 534, Protostars and Planets VII, ed. S.~{Inutsuka}, Y.~{Aikawa}, T.~{Muto}, K.~{Tomida}, \& M.~{Tamura}, 605

\bibitem[{{Benisty} {et~al.}(2017){Benisty}, {Stolker}, {Pohl}, {de Boer}, {Lesur}, {Dominik}, {Dullemond}, {Langlois}, {Min}, {Wagner}, {Henning}, {Juhasz}, {Pinilla}, {Facchini}, {Apai}, {van Boekel}, {Garufi}, {Ginski}, {M{\'e}nard}, {Pinte}, {Quanz}, {Zurlo}, {Boccaletti}, {Bonnefoy}, {Beuzit}, {Chauvin}, {Cudel}, {Desidera}, {Feldt}, {Fontanive}, {Gratton}, {Kasper}, {Lagrange}, {LeCoroller}, {Mouillet}, {Mesa}, {Sissa}, {Vigan}, {Antichi}, {Buey}, {Fusco}, {Gisler}, {Llored}, {Magnard}, {Moeller-Nilsson}, {Pragt}, {Roelfsema}, {Sauvage}, \& {Wildi}}]{Benisty2017}
{Benisty}, M., {Stolker}, T., {Pohl}, A., {et~al.} 2017, \aap, 597, A42

\bibitem[{{Beuzit} {et~al.}(2019){Beuzit}, {Vigan}, {Mouillet}, {Dohlen}, {Gratton}, {Boccaletti}, {Sauvage}, {Schmid}, {Langlois}, {Petit}, {Baruffolo}, {Feldt}, {Milli}, {Wahhaj}, {Abe}, {Anselmi}, {Antichi}, {Barette}, {Baudrand}, {Baudoz}, {Bazzon}, {Bernardi}, {Blanchard}, {Brast}, {Bruno}, {Buey}, {Carbillet}, {Carle}, {Cascone}, {Chapron}, {Charton}, {Chauvin}, {Claudi}, {Costille}, {De Caprio}, {de Boer}, {Delboulb{\'e}}, {Desidera}, {Dominik}, {Downing}, {Dupuis}, {Fabron}, {Fantinel}, {Farisato}, {Feautrier}, {Fedrigo}, {Fusco}, {Gigan}, {Ginski}, {Girard}, {Giro}, {Gisler}, {Gluck}, {Gry}, {Henning}, {Hubin}, {Hugot}, {Incorvaia}, {Jaquet}, {Kasper}, {Lagadec}, {Lagrange}, {Le Coroller}, {Le Mignant}, {Le Ruyet}, {Lessio}, {Lizon}, {Llored}, {Lundin}, {Madec}, {Magnard}, {Marteaud}, {Martinez}, {Maurel}, {M{\'e}nard}, {Mesa}, {M{\"o}ller-Nilsson}, {Moulin}, {Moutou}, {Orign{\'e}}, {Parisot}, {Pavlov}, {Perret}, {Pragt}, {Puget}, {Rabou}, {Ramos}, {Reess}, {Rigal}, {Rochat}, {Roelfsema}, {Rousset},
  {Roux}, {Saisse}, {Salasnich}, {Santambrogio}, {Scuderi}, {Segransan}, {Sevin}, {Siebenmorgen}, {Soenke}, {Stadler}, {Suarez}, {Tiph{\`e}ne}, {Turatto}, {Udry}, {Vakili}, {Waters}, {Weber}, {Wildi}, {Zins}, \& {Zurlo}}]{Beuzit2019}
{Beuzit}, J.~L., {Vigan}, A., {Mouillet}, D., {et~al.} 2019, \aap, 631, A155

\bibitem[{{Birnstiel}(2024)}]{Birnstiel2024}
{Birnstiel}, T. 2024, \araa, 62, 157

\bibitem[{{Bohren} \& {Huffman}(1983)}]{Bohren1983}
{Bohren}, C.~F. \& {Huffman}, D.~R. 1983, {Absorption and scattering of light by small particles}

\bibitem[{{Carmona} {et~al.}(2014){Carmona}, {Pinte}, {Thi}, {Benisty}, {M{\'e}nard}, {Grady}, {Kamp}, {Woitke}, {Olofsson}, {Roberge}, {Brittain}, {Duch{\^e}ne}, {Meeus}, {Martin-Za{\"\i}di}, {Dent}, {Le Bouquin}, \& {Berger}}]{Carmona2014}
{Carmona}, A., {Pinte}, C., {Thi}, W.~F., {et~al.} 2014, \aap, 567, A51

\bibitem[{{Collins} {et~al.}(2009){Collins}, {Grady}, {Hamaguchi}, {Wisniewski}, {Brittain}, {Sitko}, {Carpenter}, {Williams}, {Mathews}, {Williger}, {van Boekel}, {Carmona}, {Henning}, {van den Ancker}, {Meeus}, {Chen}, {Petre}, \& {Woodgate}}]{Collins2009}
{Collins}, K.~A., {Grady}, C.~A., {Hamaguchi}, K., {et~al.} 2009, \apj, 697, 557

\bibitem[{{Cuello} {et~al.}(2019){Cuello}, {Montesinos}, {Stammler}, {Louvet}, \& {Cuadra}}]{Cuello2019}
{Cuello}, N., {Montesinos}, M., {Stammler}, S.~M., {Louvet}, F., \& {Cuadra}, J. 2019, \aap, 622, A43

\bibitem[{{Cutri} {et~al.}(2003){Cutri}, {Skrutskie}, {van Dyk}, {Beichman}, {Carpenter}, {Chester}, {Cambresy}, {Evans}, {Fowler}, {Gizis}, {Howard}, {Huchra}, {Jarrett}, {Kopan}, {Kirkpatrick}, {Light}, {Marsh}, {McCallon}, {Schneider}, {Stiening}, {Sykes}, {Weinberg}, {Wheaton}, {Wheelock}, \& {Zacarias}}]{2003Cat}
{Cutri}, R.~M., {Skrutskie}, M.~F., {van Dyk}, S., {et~al.} 2003, {VizieR Online Data Catalog: 2MASS All-Sky Catalog of Point Sources (Cutri+ 2003)}, VizieR On-line Data Catalog: II/246. Originally published in: University of Massachusetts and Infrared Processing and Analysis Center, (IPAC/California Institute of Technology) (2003)

\bibitem[{{Dipierro} {et~al.}(2015){Dipierro}, {Pinilla}, {Lodato}, \& {Testi}}]{Dipierro2015}
{Dipierro}, G., {Pinilla}, P., {Lodato}, G., \& {Testi}, L. 2015, \mnras, 451, 974

\bibitem[{{Dullemond} {et~al.}(2012){Dullemond}, {Juhasz}, {Pohl}, {Sereshti}, {Shetty}, {Peters}, {Commercon}, \& {Flock}}]{Dullemond2012}
{Dullemond}, C.~P., {Juhasz}, A., {Pohl}, A., {et~al.} 2012, {RADMC-3D: A multi-purpose radiative transfer tool}, Astrophysics Source Code Library, record ascl:1202.015

\bibitem[{{ESA}(1997)}]{ESA1997}
{ESA}, ed. 1997, ESA Special Publication, Vol. 1200, {The HIPPARCOS and TYCHO catalogues. Astrometric and photometric star catalogues derived from the ESA HIPPARCOS Space Astrometry Mission}

\bibitem[{{Foreman-Mackey} {et~al.}(2013){Foreman-Mackey}, {Hogg}, {Lang}, \& {Goodman}}]{Foreman-Mackey2013}
{Foreman-Mackey}, D., {Hogg}, D.~W., {Lang}, D., \& {Goodman}, J. 2013, \pasp, 125, 306

\bibitem[{{Gaia Collaboration} {et~al.}(2023){Gaia Collaboration}, {Vallenari}, {Brown}, {Prusti}, {de Bruijne}, {Arenou}, {Babusiaux}, {Biermann}, {Creevey}, {Ducourant}, {Evans}, {Eyer}, {Guerra}, {Hutton}, {Jordi}, {Klioner}, {Lammers}, {Lindegren}, {Luri}, {Mignard}, {Panem}, {Pourbaix}, {Randich}, {Sartoretti}, {Soubiran}, {Tanga}, {Walton}, {Bailer-Jones}, {Bastian}, {Drimmel}, {Jansen}, {Katz}, {Lattanzi}, {van Leeuwen}, {Bakker}, {Cacciari}, {Casta{\~n}eda}, {De Angeli}, {Fabricius}, {Fouesneau}, {Fr{\'e}mat}, {Galluccio}, {Guerrier}, {Heiter}, {Masana}, {Messineo}, {Mowlavi}, {Nicolas}, {Nienartowicz}, {Pailler}, {Panuzzo}, {Riclet}, {Roux}, {Seabroke}, {Sordo}, {Th{\'e}venin}, {Gracia-Abril}, {Portell}, {Teyssier}, {Altmann}, {Andrae}, {Audard}, {Bellas-Velidis}, {Benson}, {Berthier}, {Blomme}, {Burgess}, {Busonero}, {Busso}, {C{\'a}novas}, {Carry}, {Cellino}, {Cheek}, {Clementini}, {Damerdji}, {Davidson}, {de Teodoro}, {Nu{\~n}ez Campos}, {Delchambre}, {Dell'Oro}, {Esquej},
  {Fern{\'a}ndez-Hern{\'a}ndez}, {Fraile}, {Garabato}, {Garc{\'\i}a-Lario}, {Gosset}, {Haigron}, {Halbwachs}, {Hambly}, {Harrison}, {Hern{\'a}ndez}, {Hestroffer}, {Hodgkin}, {Holl}, {Jan{\ss}en}, {Jevardat de Fombelle}, {Jordan}, {Krone-Martins}, {Lanzafame}, {L{\"o}ffler}, {Marchal}, {Marrese}, {Moitinho}, {Muinonen}, {Osborne}, {Pancino}, {Pauwels}, {Recio-Blanco}, {Reyl{\'e}}, {Riello}, {Rimoldini}, {Roegiers}, {Rybizki}, {Sarro}, {Siopis}, {Smith}, {Sozzetti}, {Utrilla}, {van Leeuwen}, {Abbas}, {{\'A}brah{\'a}m}, {Abreu Aramburu}, {Aerts}, {Aguado}, {Ajaj}, {Aldea-Montero}, {Altavilla}, {{\'A}lvarez}, {Alves}, {Anders}, {Anderson}, {Anglada Varela}, {Antoja}, {Baines}, {Baker}, {Balaguer-N{\'u}{\~n}ez}, {Balbinot}, {Balog}, {Barache}, {Barbato}, {Barros}, {Barstow}, {Bartolom{\'e}}, {Bassilana}, {Bauchet}, {Becciani}, {Bellazzini}, {Berihuete}, {Bernet}, {Bertone}, {Bianchi}, {Binnenfeld}, {Blanco-Cuaresma}, {Blazere}, {Boch}, {Bombrun}, {Bossini}, {Bouquillon}, {Bragaglia}, {Bramante}, {Breedt},
  {Bressan}, {Brouillet}, {Brugaletta}, {Bucciarelli}, {Burlacu}, {Butkevich}, {Buzzi}, {Caffau}, {Cancelliere}, {Cantat-Gaudin}, {Carballo}, {Carlucci}, {Carnerero}, {Carrasco}, {Casamiquela}, {Castellani}, {Castro-Ginard}, {Chaoul}, {Charlot}, {Chemin}, {Chiaramida}, {Chiavassa}, {Chornay}, {Comoretto}, {Contursi}, {Cooper}, {Cornez}, {Cowell}, {Crifo}, {Cropper}, {Crosta}, {Crowley}, {Dafonte}, {Dapergolas}, {David}, {David}, {de Laverny}, {De Luise}, \& {De March}}]{Gaia2023}
{Gaia Collaboration}, {Vallenari}, A., {Brown}, A.~G.~A., {et~al.} 2023, \aap, 674, A1

\bibitem[{{Garcia} \& {Gonzalez}(2020)}]{Garcia2020}
{Garcia}, A. J.~L. \& {Gonzalez}, J.-F. 2020, \mnras, 493, 1788

\bibitem[{{Gonzalez} {et~al.}(2020){Gonzalez}, {van der Plas}, {Pinte}, {Cuello}, {Nealon}, {M{\'e}nard}, {Revol}, {Rodet}, {Langlois}, \& {Maire}}]{Gonzalez2020}
{Gonzalez}, J.-F., {van der Plas}, G., {Pinte}, C., {et~al.} 2020, \mnras, 499, 3837

\bibitem[{{Henyey} \& {Greenstein}(1941)}]{Henyey1941}
{Henyey}, L.~G. \& {Greenstein}, J.~L. 1941, \apj, 93, 70

\bibitem[{{H{\o}g} {et~al.}(2000){H{\o}g}, {Fabricius}, {Makarov}, {Urban}, {Corbin}, {Wycoff}, {Bastian}, {Schwekendiek}, \& {Wicenec}}]{2000Cat}
{H{\o}g}, E., {Fabricius}, C., {Makarov}, V.~V., {et~al.} 2000, \aap, 355, L27

\bibitem[{{Hughes} {et~al.}(2018){Hughes}, {Duch{\^e}ne}, \& {Matthews}}]{Hughes2018}
{Hughes}, A.~M., {Duch{\^e}ne}, G., \& {Matthews}, B.~C. 2018, \araa, 56, 541

\bibitem[{{Khalafinejad} {et~al.}(2016){Khalafinejad}, {Maaskant}, {Mari{\~n}as}, \& {Tielens}}]{Khalafinejad2016}
{Khalafinejad}, S., {Maaskant}, K.~M., {Mari{\~n}as}, N., \& {Tielens}, A.~G.~G.~M. 2016, \aap, 587, A62

\bibitem[{{Long} {et~al.}(2017){Long}, {Fernandes}, {Sitko}, {Wagner}, {Muto}, {Hashimoto}, {Follette}, {Grady}, {Fukagawa}, {Hasegawa}, {Kluska}, {Kraus}, {Mayama}, {McElwain}, {Oh}, {Tamura}, {Uyama}, {Wisniewski}, \& {Yang}}]{Long2017}
{Long}, Z.~C., {Fernandes}, R.~B., {Sitko}, M., {et~al.} 2017, \apj, 838, 62

\bibitem[{{Ma} \& {Schmid}(2022)}]{Ma2022}
{Ma}, J. \& {Schmid}, H.~M. 2022, \aap, 663, A110

\bibitem[{{Ma} {et~al.}(2024){Ma}, {Schmid}, \& {Stolker}}]{Ma2024}
{Ma}, J., {Schmid}, H.~M., \& {Stolker}, T. 2024, \aap, 683, A18

\bibitem[{{Malfait} {et~al.}(1998){Malfait}, {Bogaert}, \& {Waelkens}}]{Malfait1998}
{Malfait}, K., {Bogaert}, E., \& {Waelkens}, C. 1998, \aap, 331, 211

\bibitem[{{Meeus} {et~al.}(2001){Meeus}, {Waters}, {Bouwman}, {van den Ancker}, {Waelkens}, \& {Malfait}}]{Meeus2001}
{Meeus}, G., {Waters}, L.~B.~F.~M., {Bouwman}, J., {et~al.} 2001, \aap, 365, 476

\bibitem[{{Michoulier} {et~al.}(2024){Michoulier}, {Gonzalez}, \& {Price}}]{Michoulier2024}
{Michoulier}, S., {Gonzalez}, J.-F., \& {Price}, D.~J. 2024, \aap, 688, A31

\bibitem[{{Min} {et~al.}(2009){Min}, {Dullemond}, {Dominik}, {de Koter}, \& {Hovenier}}]{Min2009}
{Min}, M., {Dullemond}, C.~P., {Dominik}, C., {de Koter}, A., \& {Hovenier}, J.~W. 2009, \aap, 497, 155

\bibitem[{{Min} {et~al.}(2005){Min}, {Hovenier}, \& {de Koter}}]{Min2005}
{Min}, M., {Hovenier}, J.~W., \& {de Koter}, A. 2005, \aap, 432, 909

\bibitem[{{Min} {et~al.}(2016){Min}, {Rab}, {Woitke}, {Dominik}, \& {M{\'e}nard}}]{Min2016}
{Min}, M., {Rab}, C., {Woitke}, P., {Dominik}, C., \& {M{\'e}nard}, F. 2016, \aap, 585, A13

\bibitem[{{Mulders} {et~al.}(2013){Mulders}, {Min}, {Dominik}, {Debes}, \& {Schneider}}]{Mulders2013}
{Mulders}, G.~D., {Min}, M., {Dominik}, C., {Debes}, J.~H., \& {Schneider}, G. 2013, \aap, 549, A112

\bibitem[{{Nealon} {et~al.}(2020){Nealon}, {Cuello}, {Gonzalez}, {van der Plas}, {Pinte}, {Alexander}, {M{\'e}nard}, \& {Price}}]{Nealon2020}
{Nealon}, R., {Cuello}, N., {Gonzalez}, J.-F., {et~al.} 2020, \mnras, 499, 3857

\bibitem[{{Pinilla} {et~al.}(2016){Pinilla}, {Klarmann}, {Birnstiel}, {Benisty}, {Dominik}, \& {Dullemond}}]{Pinilla2016}
{Pinilla}, P., {Klarmann}, L., {Birnstiel}, T., {et~al.} 2016, \aap, 585, A35

\bibitem[{{Ren} {et~al.}(2020){Ren}, {Pueyo}, {Chen}, {Choquet}, {Debes}, {Duch{\^e}ne}, {M{\'e}nard}, \& {Perrin}}]{Ren2020}
{Ren}, B., {Pueyo}, L., {Chen}, C., {et~al.} 2020, \apj, 892, 74

\bibitem[{{Ren} {et~al.}(2023){Ren}, {Benisty}, {Ginski}, {Tazaki}, {Wallack}, {Milli}, {Garufi}, {Bae}, {Facchini}, {M{\'e}nard}, {Pinilla}, {Swastik}, {Teague}, \& {Wahhaj}}]{Ren2023}
{Ren}, B.~B., {Benisty}, M., {Ginski}, C., {et~al.} 2023, \aap, 680, A114

\bibitem[{{Rosotti} {et~al.}(2020){Rosotti}, {Benisty}, {Juh{\'a}sz}, {Teague}, {Clarke}, {Dominik}, {Dullemond}, {Klaassen}, {Matr{\`a}}, \& {Stolker}}]{Rosotti2020}
{Rosotti}, G.~P., {Benisty}, M., {Juh{\'a}sz}, A., {et~al.} 2020, \mnras, 491, 1335

\bibitem[{{Schmid} {et~al.}(2018){Schmid}, {Bazzon}, {Roelfsema}, {Mouillet}, {Milli}, {Menard}, {Gisler}, {Hunziker}, {Pragt}, {Dominik}, {Boccaletti}, {Ginski}, {Abe}, {Antoniucci}, {Avenhaus}, {Baruffolo}, {Baudoz}, {Beuzit}, {Carbillet}, {Chauvin}, {Claudi}, {Costille}, {Daban}, {de Haan}, {Desidera}, {Dohlen}, {Downing}, {Elswijk}, {Engler}, {Feldt}, {Fusco}, {Girard}, {Gratton}, {Hanenburg}, {Henning}, {Hubin}, {Joos}, {Kasper}, {Keller}, {Langlois}, {Lagadec}, {Martinez}, {Mulder}, {Pavlov}, {Podio}, {Puget}, {Quanz}, {Rigal}, {Salasnich}, {Sauvage}, {Schuil}, {Siebenmorgen}, {Sissa}, {Snik}, {Suarez}, {Thalmann}, {Turatto}, {Udry}, {van Duin}, {van Holstein}, {Vigan}, \& {Wildi}}]{Schmid2018}
{Schmid}, H.~M., {Bazzon}, A., {Roelfsema}, R., {et~al.} 2018, \aap, 619, A9

\bibitem[{{Schmid} \& {Ma}(2025)}]{Schmid2025}
{Schmid}, H.~M. \& {Ma}, J. 2025, submitted to A\&A

\bibitem[{{Soummer} {et~al.}(2012){Soummer}, {Pueyo}, \& {Larkin}}]{Soummer2012}
{Soummer}, R., {Pueyo}, L., \& {Larkin}, J. 2012, \apjl, 755, L28

\bibitem[{{Tanaka} {et~al.}(2023){Tanaka}, {Anayama}, \& {Tazaki}}]{Tanaka2023}
{Tanaka}, H., {Anayama}, R., \& {Tazaki}, R. 2023, \apj, 945, 68

\bibitem[{{Tatulli} {et~al.}(2011){Tatulli}, {Benisty}, {M{\'e}nard}, {Varni{\`e}re}, {Martin-Za{\"\i}di}, {Thi}, {Pinte}, {Massi}, {Weigelt}, {Hofmann}, \& {Petrov}}]{Tatulli2011}
{Tatulli}, E., {Benisty}, M., {M{\'e}nard}, F., {et~al.} 2011, \aap, 531, A1

\bibitem[{{Tazaki} \& {Dominik}(2022)}]{Tazaki2022}
{Tazaki}, R. \& {Dominik}, C. 2022, \aap, 663, A57

\bibitem[{{Tazaki} {et~al.}(2023){Tazaki}, {Ginski}, \& {Dominik}}]{Tazaki2023}
{Tazaki}, R., {Ginski}, C., \& {Dominik}, C. 2023, \apjl, 944, L43

\bibitem[{{Tazaki} {et~al.}(2019){Tazaki}, {Tanaka}, {Muto}, {Kataoka}, \& {Okuzumi}}]{Tazaki2019}
{Tazaki}, R., {Tanaka}, H., {Muto}, T., {Kataoka}, A., \& {Okuzumi}, S. 2019, \mnras, 485, 4951

\bibitem[{{Tschudi} \& {Schmid}(2021)}]{Tschudi2021}
{Tschudi}, C. \& {Schmid}, H.~M. 2021, \aap, 655, A37

\bibitem[{{van der Plas} {et~al.}(2019){van der Plas}, {M{\'e}nard}, {Gonzalez}, {Perez}, {Rodet}, {Pinte}, {Cieza}, {Casassus}, \& {Benisty}}]{vanderplas2019}
{van der Plas}, G., {M{\'e}nard}, F., {Gonzalez}, J.~F., {et~al.} 2019, \aap, 624, A33

\bibitem[{{van Holstein} {et~al.}(2020){van Holstein}, {Girard}, {de Boer}, {Snik}, {Milli}, {Stam}, {Ginski}, {Mouillet}, {Wahhaj}, {Schmid}, {Keller}, {Langlois}, {Dohlen}, {Vigan}, {Pohl}, {Carbillet}, {Fantinel}, {Maurel}, {Orign{\'e}}, {Petit}, {Rigal}, {Sevin}, {Boccaletti}, {Le Coroller}, {Dominik}, {Henning}, {Lagadec}, {M{\'e}nard}, {Turatto}, {Udry}, {Chauvin}, {Feldt}, \& {Beuzit}}]{vanHolstein2020}
{van Holstein}, R.~G., {Girard}, J.~H., {de Boer}, J., {et~al.} 2020, {IRDAP: SPHERE-IRDIS polarimetric data reduction pipeline}, Astrophysics Source Code Library, record ascl:2004.015

\bibitem[{{Wagner} {et~al.}(2015){Wagner}, {Apai}, {Kasper}, \& {Robberto}}]{Wagner2015}
{Wagner}, K., {Apai}, D., {Kasper}, M., \& {Robberto}, M. 2015, \apjl, 813, L2

\bibitem[{{Wagner} {et~al.}(2018){Wagner}, {Dong}, {Sheehan}, {Apai}, {Kasper}, {McClure}, {Morzinski}, {Close}, {Males}, {Hinz}, {Quanz}, \& {Fung}}]{Wagner2018}
{Wagner}, K., {Dong}, R., {Sheehan}, P., {et~al.} 2018, \apj, 854, 130

\bibitem[{{Wahhaj} {et~al.}(2021){Wahhaj}, {Milli}, {Romero}, {Cieza}, {Zurlo}, {Vigan}, {Pe{\~n}a}, {Valdes}, {Cantalloube}, {Girard}, \& {Pantoja}}]{Wahhaj2021}
{Wahhaj}, Z., {Milli}, J., {Romero}, C., {et~al.} 2021, \aap, 648, A26

\bibitem[{{Woitke} {et~al.}(2016){Woitke}, {Min}, {Pinte}, {Thi}, {Kamp}, {Rab}, {Anthonioz}, {Antonellini}, {Baldovin-Saavedra}, {Carmona}, {Dominik}, {Dionatos}, {Greaves}, {G{\"u}del}, {Ilee}, {Liebhart}, {M{\'e}nard}, {Rigon}, {Waters}, {Aresu}, {Meijerink}, \& {Spaans}}]{Woitke2016}
{Woitke}, P., {Min}, M., {Pinte}, C., {et~al.} 2016, \aap, 586, A103

\bibitem[{{Xie} {et~al.}(2023){Xie}, {Ren}, {Dong}, {Choquet}, {Vigan}, {Gonzalez}, {Wagner}, {Fang}, \& {Ubeira-Gabellini}}]{Xie2023}
{Xie}, C., {Ren}, B.~B., {Dong}, R., {et~al.} 2023, \aap, 675, L1

\bibitem[{{Zhu} {et~al.}(2012){Zhu}, {Nelson}, {Dong}, {Espaillat}, \& {Hartmann}}]{Zhu2012}
{Zhu}, Z., {Nelson}, R.~P., {Dong}, R., {Espaillat}, C., \& {Hartmann}, L. 2012, \apj, 755, 6

\end{thebibliography}

\appendix
\section{Photometric calibrations}
\label{app:calib}

\subsection{Background subtraction}
\label{app: bck-subtr}
\begin{figure}
    \centering
    \includegraphics[width=0.49\textwidth]{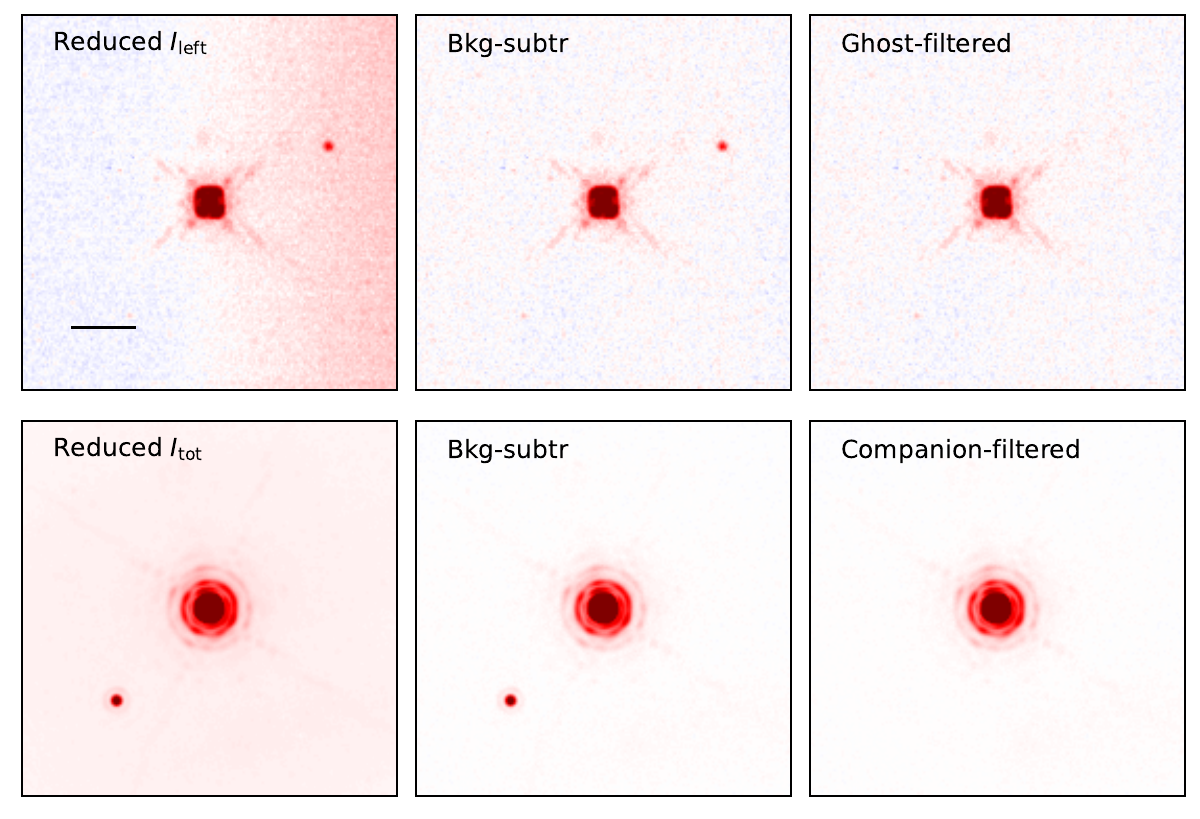}
    \vspace{-0.3cm}
    \caption{Example of background subtraction and companion/ghost filtration. Top left: $I_{\rm left}$ image of flux frame taken before all polarimetric cycles in K-hop. Top middle: $I_{\rm left}$ with the background subtracted. Top right:  $I_{\rm left}$ with the background subtracted and with the ghost filtered. 
    Bottom left: $I_{\rm tot}$ of the first cycle from the K sequence. Bottom middle: $I_{\rm tot}$ with the background subtracted. Bottom right: $I_{\rm tot}$ with the background subtracted and the companion filtered. }
    \label{fig: bck}
\end{figure}

The thermal background noise is known to be an issue for $Ks$ band observation \citep{Beuzit2019}, not important at shorter wavelengths. Therefore, we applied background subtraction to $Ks$ band PSF images. 
For K-hop sequence, sky images were taken only for saturated object frames but not for flux frames for Ks deep observing sequences. Therefore, although the pipeline estimates and subtracts a flat background signal in a large annulus, the lack of sky images still results in a background in the reduced images, see Fig.~\ref{fig: bck} top left image. In addition, a faint ghost is seen, while the companion is not visible in this short exposure. Similarly, no sky images were taken for K-sequence, which results in nonnegligeable background signal, as shown in Fig.~\ref{fig: bck} bottom left image. The companion is well-visible in the SE direction. 

For K-hop, we performed a second-order polynomial fitting to the background in the two $I_{\rm left}$ and $I_{\rm right}$ images. The long-exposures have sky-frames so $Q$ and $U$ are also not affected.  
For K sequence, we performed a linear fitting to the background in the radial direction for each cycle of $I_{\rm tot}$ images. The background is at the same level for all frames so $Q$ and $U$ images are not affected. The background-subtracted images are shown in the middle column of Fig.~\ref{fig: bck}. This step is important for an accurate quantitative measurement of integrated stellar flux $I_{\star}$. 

After the background subtraction, we filtered the ghost on the west for K-hop and the companion for K. The flux frames after the background subtraction and ghost/companion filtering are present on the right columns of Fig.~\ref{fig: bck}. 

\subsection{Coronagraph transmission correction}
\label{app: coronagraph-correction}
\begin{figure}
    \centering
    \includegraphics[width=\linewidth]{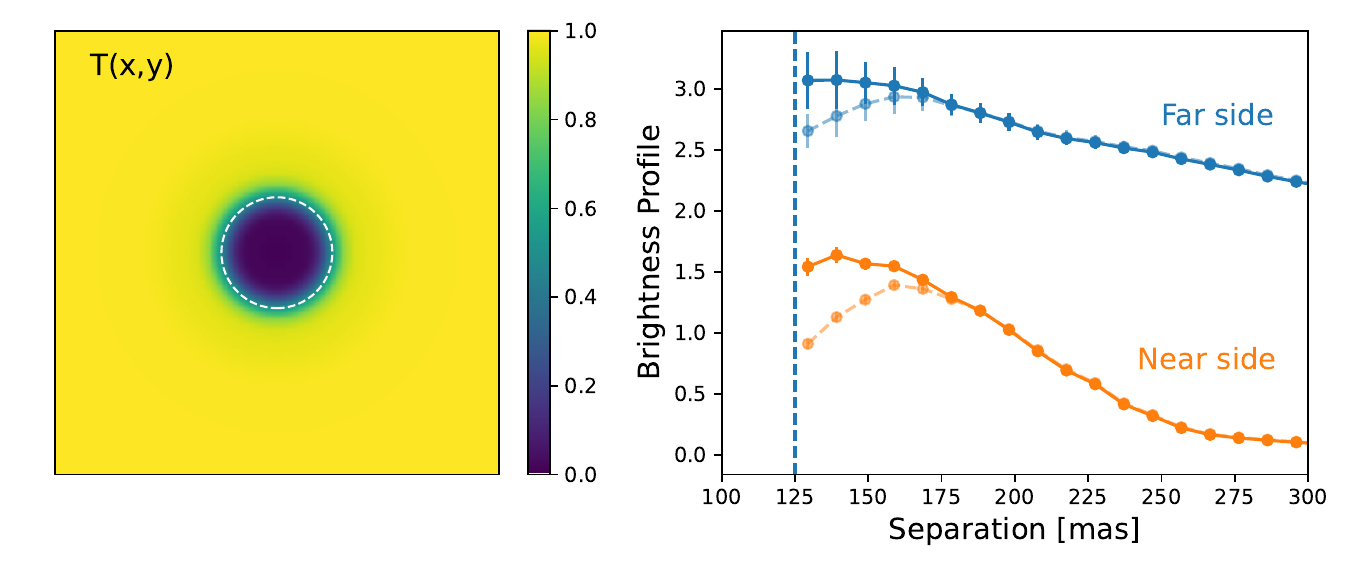}
    \caption{Coronagraph N\_ALC\_Ks transmission correction. Left: Transmission map $T(x,y)$ of coronagraph N\_ALC\_Ks zoomed to 1$\times$1\arcsec. Right: Radial profile of the total disk intensity $I_{\rm disk}(r)$ for the near and far sides. The dashed and solid lines represent the profiles before and after the correction respectively. }
    \label{fig:cor_alc}
\end{figure}

The coronagraph N\_ALC\_Ks combines the apodizer and the Lyot stop of diameter 0.24\arcsec. Although the suggested inner working angle (IWA) in the manual is  $r=0.2\arcsec$, we tried to recover the intensity close to 0.125\arcsec using the radial transmission curve\footnote{See section A.6 from SPHERE user manual at \url{https://www.eso.org/sci/facilities/paranal/instruments/sphere/doc/VLT-MAN-SPH-14690-0430_P114_july_2024_zwa.pdf }}. We interpolated the curve and corrected the flux between 0.125\arcsec and 0.5\arcsec, as shown in Fig.~\ref{fig:cor_alc} left panel. We consider the IWA = 0.125\arcsec at which the transmission is 50\%, indicated in Fig.~\ref{fig:cor_alc} as the white dashed line, and discard the data inward. Then for each image frame, we corrected the transmission by dividing $T(x,y)$ pixel-wise. In the right panel, we show the radial profiles of the total intensity $I_{\rm disk}(r)$ on the near and far sides. The profiles are normalized to the peak value on the major axis, and the far side is shifted by 2 for clarity.

This correction is important for HD~100453 not only for integrated flux but also for brightness ratios.  Outside the IWA, the correction is well characterized and improves the flux estimates.  Without applying transmission correction, the integrated flux $I_{\rm disk}/I_{\star} = 3.5\%$ and $\hat{Q}_{\varphi}/I_{\star} = 1.18\%$, meaning $20\%$ and $6\%$ flux underestimation in $I_{\rm disk}$ and $Q_{\varphi}$ respectively. 
Since the main correction is on the near and far side of the disk, and near side loses more flux as it is closer to the center from projection effect, the $R_{N/F}$ and $R_{N/M}$ will be underestimated in both the polarized intensity and total intensity if not corrected. The coronagraph also pushes the peak intensity outward or even creates an artificial peak. 

\subsection{Flux of companion}
\label{app: companion}
To estimate the relative flux ratio between the host star and the companion, we performed aperture photometry using circular aperture $r=0.1\arcsec$ with local background estimated in the surrounding annulus $r = [0.10, 0.11]\arcsec$.
We integrate the flux for the companion $I^{s}_{B}$ and for the star $I^{s}_{A}$ for each cycle. 
Since the companion is too faint in $V$-band short exposures, we do not have simultaneous unsaturated A and B at this wavelength. We measure $I^{s}_{B}$ in long exposures and $I^{s}_{A}$ in short exposures separately. For $I$, $J$, and $Ks$, unsaturated star A and B are clearly visible in each total intensity frame, and we calculate the flux ratio $I^{s}_{B}/I^{s}_{A}$cycle by cycle. The results are summarized in Table.~\ref{tab:f_companion}. 

Then we adopt the magnitude $m_{A}$ for HD~100453 A\citep{2000Cat, 2003Cat, ESA1997} and derive the magnitude of HD~100453 B, including the first magnitude value at $J$-band. The $m_{B}$ at $V$ agree with previous studies in \citet{Collins2009}, which measured $m_{B}=15.88$ using HST F606W image. We did not include $m_{B}$ at $Ks$ because the inner disk largely contributes to the central radiation at $Ks$ \citep{Khalafinejad2016}, and the disk intensity might be variable making $I_{A}$ at this wavelength not a reliable reference to determine $m_{B}$. 



\begin{table}[]
    \centering
    \caption{Flux and magnitude of the companion HD~100453B.}
    \begin{tabular}{c c c c}
    \hline
    \hline
    Filter  &  $I^{s}_{B}/I^{s}_{A}$[\%] & $m_{A}$ & $m_{B}$ \\
    \hline
     V      &  0.054 $\pm$ 0.008 & 7.79$\pm$0.01 & 15.9 $\pm$ 0.2\\
     I      &  0.43 $\pm$ 0.02 & 7.41$\pm$0.20 & 13.3 $\pm$ 0.2 \\
     J      &  1.60 $\pm$ 0.04 & 6.94$\pm$0.03 & 11.43 $\pm$ 0.04\\
     K      &  1.28 $\pm$ 0.01 & - & -  \\
    \hline
    \end{tabular}
    \label{tab:f_companion}
\end{table}

\section{Disk intensity extraction}
\label{app:disk-intensity}
\subsection{Halo component}
Including the halo component in RDI is necessary for $V$ band. To first-order approximation, the scaling factor in Eq.~\ref{eq: disk-intensity} can be estimated as $a\approx \Sigma{I_{\star}}/\Sigma{I_{\rm ref}}$, where $\Sigma$ denotes integration over a disk-free region. However, if the selected fitting region is contaminated by extended disk signal (i.e., the halo component $I_{\rm halo}$), which is often the case, the effective scaling becomes $a'\approx \Sigma{(I_{\star}+I_{\rm halo})}/{\Sigma{I_{\rm ref}}}>a$, thus overestimating the scaling factor by $\Delta a = \Sigma{I_{\rm halo}}/\Sigma{I_{\rm ref}}$. 

When applied to the subtraction on the disk region, this leads to oversubtraction of the disk signal at each pixel $\Delta{I_{\rm disk}}(x,y) = \Delta a \cdot I_{\rm ref}(x,y)$. This oversubtraction becomes significant when the relative error, 
\begin{equation}
    \dfrac{\Delta{I_{\rm disk}}}{I_{\rm disk}}(x,y) = \dfrac{\Sigma{I_{\rm halo}}}{\Sigma{I_{\rm ref}}}\cdot \dfrac{I_{\rm ref}}{I_{\rm disk}}(x,y),
\end{equation}
is nonnegligible. Thus, even a small amount of halo contamination can lead to large errors if the disk signal is faint relative to the reference.

Using the Setup A model at $V$ band, we calculated the radial profile $I_{\rm disk}(r)$ and fitted the halo profile in $r=[0.43, 0.76]\arcsec$ with an exponential drop-off, resulting in $I_{\rm disk}(r) \propto e^{-4.5\ r}$. 

We estimated $\Sigma{I_{\rm halo}}/\Sigma{I_{\rm ref}} \approx 1\%$ in the chosen fitting region and $I_{\rm ref}/I_{\rm disk}(x,y) \approx 20 $ at the disk intensity peak location. This implies that neglecting the halo component leads to about $20\%$ underestimation of the $I_{\rm disk}(x,y)$ at peak position. The extracted disk intensity ignoring this halo component is plot in the left panel of Fig.~\ref{fig:mask-r-k}, clearly showing the oversubstraction. 

For comparison, we applied the same estimation to $Ks$-band. We find $\Sigma{I_{\rm halo}}/\Sigma{I_{\rm ref}} \approx 4\%$ in the chosen fitting region, but $I_{\rm ref/I_{\rm disk}}(x,y) \approx 0.5 $ at the peak of the disk because the disk is brighter and the star is fainter at longer wavelength. This results in only a $2\%$ underestimation of the $I_{\rm disk}$. Therefore, we consider the contribution of $I_{\rm halo}$ as negligible at the $Ks$-band.  

\subsection{Excluding companion}
\label{app:nmf-imaging}
Although the $Ks$ band disk intensity extraction does not suffer from the extended disk halo created by the convolution effect, including the companion into the fitting region results in a similar effect. 
This effect is seen when we compare our results with that from \citet{Ren2023}, where a faint negative halo present because of including the companion into the fitting region. Therefore, the disk intensity is oversubtracted, and the disk signal is underestimated.
We followed the steps as described in \citet{Ren2023}, but using the imputation mask excluding the companion (indicated by the gray shaded region in Fig.~\ref{fig:mask-r-k}). We also applied the transmission correction as described in Appendix.~\ref{app: coronagraph-correction} and the result is shown in Fig.~\ref{fig:mask-r-k} right panel. We see that the improved data imputation result is comparable to the classic RDI (shown in Fig.~\ref{fig:disk-intensity-vk}), integrating to $I_{\rm disk}/I_{\star} = 3.86\%$ agrees well with our results using classic RDI.

In summary, we conclude that false-flagging pixels containing disk signal or any other contamination in the fitting region or the data imputation can be dangerous. When we adopt the same good fitting region in our case of HD100453, the classic RDI and the data imputation results agree pretty well.

\begin{figure}
    \centering
    \includegraphics[width=0.9\linewidth]{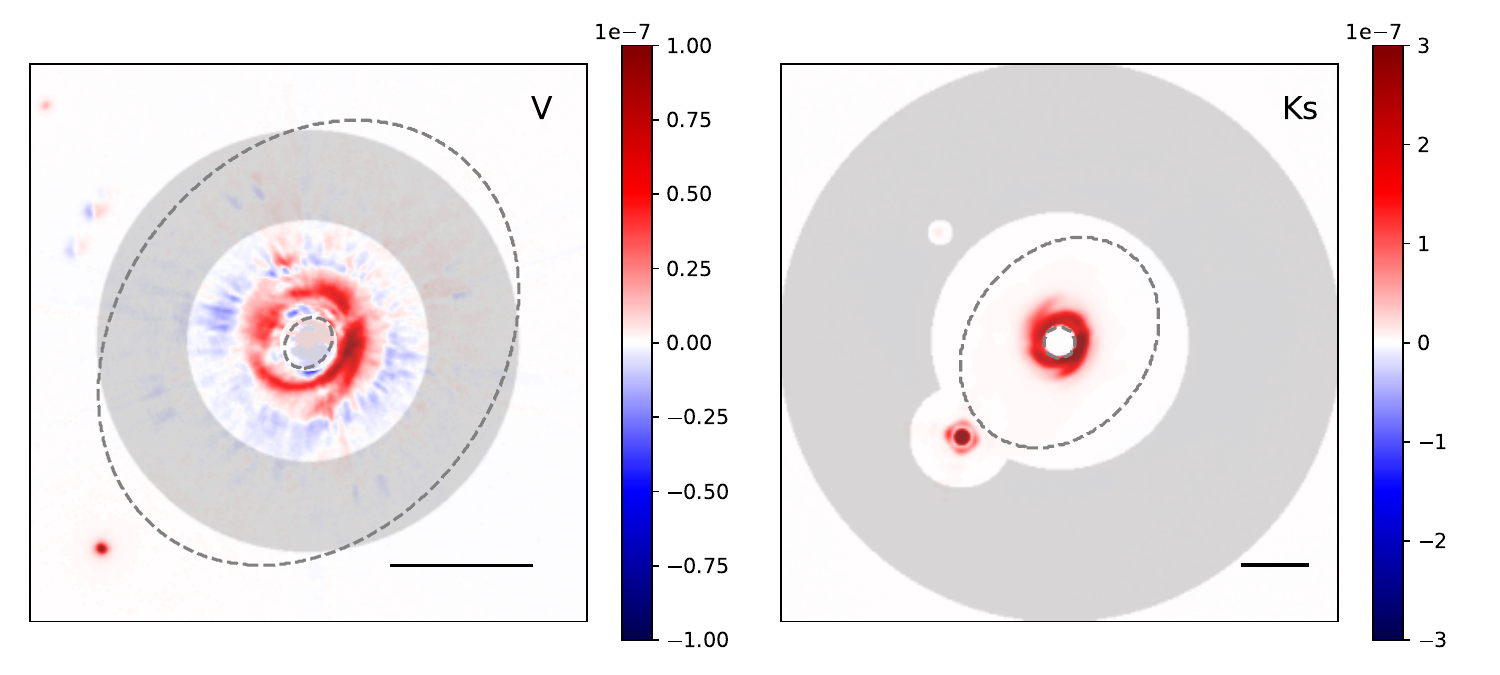}
    \caption{Fitting region, $\Sigma$, used for disk intensity extraction. Left: Fitting region at $V$-band. The gray shaded annulus indicates $\Sigma$, overplotted on $I_{\rm disk}(x,y)(V)$ extracted without introducing halo component. Right: Fitting region at $Ks$ band. The gray shaded region indicates $\Sigma$ at $Ks$-band, overplotted on $I_{\rm disk}(x,y)(K)$ extracted using NMF\_imaging with improved $\Sigma$. 
    The two images are shown in different scales to highlight the $\Sigma$, with the black bar indicating 0.5\arcsec. The gray dashed lines encloses the region for disk intensity integration. }
    \label{fig:mask-r-k}
\end{figure}

\section{Model parameters}
\begin{figure}
    \centering
    \includegraphics[width=0.8\linewidth]{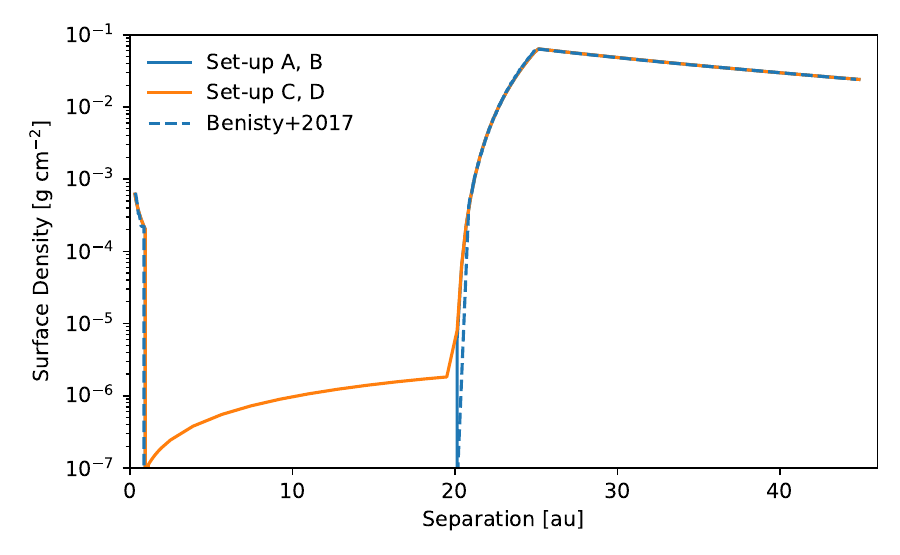}
    \caption{Surface density used in the radiative transfer model. }
    \label{fig:surface-density}
\end{figure}

The model parameters used in the RADMC-3D model are summarized in Table.~\ref{tab: model-modified}. The model is built based on the description in \citet{Benisty2017}. The main difference lies in the adjustment of spiral geometry and the new cavity zone. The spiral geometry parameters are slightly modified (highlighted in Table.~\ref{tab: model-modified}) mostly because we ignore the dust settling in this study. 
The new cavity zone only appears in Setups C and D, with its profile shown in Fig.~\ref{fig:surface-density}. 
\begin{table}
    \centering
    \caption{RADMC-3D model parameters used in this work. }
    \begin{tabular}{c c c c}
    \hline 
    \hline
        Parameter & Inner disk & Outer disk & Cavity \\
    \hline
        $R_{\rm in}$ [AU] & 0.27 & 20 & 1\\
        $R_{\rm out}$ [AU] & 1 & 48 & 20\\
        $R_{\rm tap}$ [AU] & 50 & 50 & 50 \\
        \hline
        $\epsilon$ & 1 & -3/1 & -1 \\
        $H_0/r_0$ & 0.04 & 0.05 & 0.04 \\
        $r_0$[AU] & 1 & 20 & 1 \\
        $\psi$ & 0 & 0.13 & 0\\
        \hline
        $i$ [deg] & 48 & -38 & -38 \\
        $PA $[deg] & 80 & 142 & 142 \\
        $M_{\rm dust}[M_\odot]$ & $1 \times 10^{-10}$ & $2 \times 10^{-5}$ & $1.5 \times 10^{-10}$ \\
        \hline
        Parameter & NE spiral & SW spiral \\
    \hline
        $R_{\rm in}$ [AU] & 25\tablefootmark{a} & 33\tablefootmark{a}\\
        $R_{\rm out}$ [AU] & 36\tablefootmark{a} & 43\tablefootmark{a}\\
        $A1$ & 25\tablefootmark{a} & 32\tablefootmark{a} \\
        $A2$ & 7 & 8 \\
        $\theta_0$ & 125 & 315 \\
        $n$ & 1.12 & 1.12 \\
        $a_{\rm height}$ & 0.8 & 1.2\tablefootmark{a}  \\
        $w$ [AU] & 1.2 & 1.2 \\
        q & 1.7 & 1.7 \\
    \hline
    \end{tabular}
    \tablefoot{
    \tablefoottext{a}{Parameters related to spiral geometry that are slightly modified compared with \citet{Benisty2017}. }}
    \label{tab: model-modified}
\end{table}

%
%

\end{document}